\documentclass{ar-1col}

\usepackage{amsmath,amssymb,mathtools}
\usepackage{amsfonts,bm,bbm}   
\usepackage[comma]{natbib}
\usepackage{url}
\usepackage{cite}

\jname{Annu. Rev. Fluid Mech.}
\jvol{54}
\jyear{2022}
\doi{10.1146/annurev-fluid-021021-102043}

\begin{document}

\markboth{E. Falcon \& N. Mordant}{Experiments in Surface Gravity-Capillary Wave Turbulence}

\title{Experiments in Surface Gravity-Capillary Wave Turbulence}

\author{Eric Falcon$^1$ and Nicolas Mordant$^2$
\affil{$^1$MSC (Matière et Systèmes Complexes), Universit\'e de Paris, CNRS, F-75 013 Paris, France; email: eric.falcon@u-paris.fr}
\affil{$^2$LEGI (Laboratoire des {\'E}coulements Géophysiques et Industriels), Universit\'e Grenoble Alpes, CNRS, F-38 000 Grenoble, France; email: Nicolas.Mordant@univ-grenoble-alpes.fr}}
\begin{abstract} 
The last decade has seen a significant increase in the number of studies devoted to wave turbulence. Many deal with water waves, as modeling of ocean waves has historically motivated the development of weak turbulence theory, which adresses the dynamics of a random ensemble of weakly nonlinear waves in interaction. Recent advances in experiments have shown that this theoretical picture is too idealized to capture experimental observations. While gravity dominates much of the oceanic spectrum, waves observed in the laboratory are in fact gravity-capillary waves, due to the restricted size of wave basins. This richer physics induces many interleaved physical effects far beyond the theoretical framework, notably in the vicinity of the gravity-capillary crossover. These include dissipation, finite-system size effects, and finite nonlinearity effects. Simultaneous space-and-time resolved techniques, now available, open the way for a much more advanced analysis of these effects.
\end{abstract} 

\maketitle


\section{INTRODUCTION}
Wave turbulence is generically a statistical state made of a large number of random nonlinearly coupled waves. The canonical example is ocean waves. Stormy winds excite large-scale surface gravity waves and, due to their nonlinear interaction, the wave energy is redistributed to smaller scales. This energy transfer leads ultimately to an energy cascade from the large (forcing) scale down to small (dissipative) scales. 

Weak turbulence theory was developed in the 1960s to describe the statistical properties of wave turbulence in the limits of both weak nonlinearity and an infinite system. It was initially motivated to model the ocean wave spectrum  \citep{Hasselmann62} and has since been applied in almost all domains of physics involving waves (see books by \citet{ZakharovBook1992,NazarenkoBook2011}). Despite its success at deriving a statistical theory and analytically predicting the wave spectrum in an out-of-equilibrium stationary state, weak turbulence theory does not capture all phenomena observed in nature or in the laboratory. For instance, the formation of strongly nonlinear structures (called coherent structures) such as sharp-crested waves, result from strong correlations of phases between waves, which is at odds with the theoretical hypotheses.  For a long time, measurements were mostly restricted to single-point measurements that limited the wave field analysis and specifically the detection of such coherent structures. The last decade has seen the development of measurements simultaneously resolved in space and time that shed new light on the field of wave turbulence. These have indeed opened the possibility of probing in detail the spectral content of turbulent wave fields simultaneously in wavevector $\mathbf k$ and frequency $\omega$ space, as required to discriminate propagating waves from other structures with distinct dynamics. Following these developments, experiments have explored a broader range of systems beyond water waves including vibrating plates \citep{CobelliPRL2009}, hydroelastic waves \citep{DeikeJFM2013}, inertial waves in rotating fluids \citep{MonsalvePRL2020}, and internal waves in stratified fluids \citep{SavaroPRF2020,DavisPRL2020} (this list being far from exhaustive), not to mention the numerical simulations. It is not an exaggeration to claim that the topic of wave turbulence is blooming nowadays.  

Here, we focus our review on experiments concerning gravity-capillary waves at the surface of a fluid. Laboratory experiments are restricted to wavelengths typically smaller than a meter, even in large-scale wave basins. At these scales the physics is rendered highly complex by the interplay of many physical effects. First, waves are supported either by gravity or by capillarity, with a transition at wavelengths close to the centimeter scale, at which the two contributions are deeply entangled. Second, the finite-system size effects induce the existence of discrete Fourier modes that affect the energy cascade. Third, viscous dissipation, often strongly amplified by surface contamination, also alters the energy flux cascading through the scales. Finally, the degree of nonlinearity required for the development of wave turbulence at the surface of water in experiments is not vanishingly weak as assumed in theory, and this may lead to the formation of coherent structures in addition to waves. Due to new experimental techniques developed since the turn of the twenty-first century, the interplay of all these effects can now be explored. In this article, we review the experimental works of the last two decades that address all these aspects of wave turbulence. Complementary details can be found in previous reviews on wave turbulence \citep{FalconDCDSB2010,Newell2011,NazarenkoRev2016,Hawai,Zakharov2019,GaltierAFD2020} or theoretical textbooks \citep{ZakharovBook1992,NazarenkoBook2011}. 

The review is organized as follows. We first describe briefly in Section \ref{gc} the general features of gravity-capillary surface waves, and in Section \ref{interactions} we look at the fundamental mechanism of wave turbulence, i.e., the nonlinear wave interactions. We then introduce in Section \ref{WWT} the phenomenology and the assumptions of weak turbulence theory, and we discuss in Section~\ref{Timescale} the various timescales involved in wave turbulence. We then present in Section \ref{methods} the main experimental techniques, including single-point measurements and space-and-time-resolved measurements. We discuss the main experimental results in Section \ref{1Dspectrum} and \ref{2Dspectrum} by comparing them with weak turbulence predictions and highlighting the abovementioned effects not taken into account at the current stage of theoretical developments. Finally, we present in Section \ref{largescales} a short discussion of large-scale properties in wave turbulence before ending with lists of Summary Points and Future Issues.

\section{GRAVITY-CAPILLARY DISPERSION RELATION}\label{gc}
The linear dispersion relation of inviscid deep-water waves is \citep{Lamb1932}
\begin{equation}
\omega=\sqrt{gk+\frac{\gamma}{\rho}k^3}  {\rm ,}
\label{RD}
\end{equation}
with $\omega=2\pi f$ the angular frequency, $k=2\pi/\lambda$ the wavenumber, $g$ the acceleration of gravity, $\rho$ the density of the liquid, and $\gamma$ the surface tension. 
The crossover between the gravity wave regime and the capillary wave regime occurs for a wavelength $\lambda _{gc}=2\pi \sqrt{\gamma/(\rho g)}$ close to the centimeter for most fluids. Weak turbulence theory is developed for systems with a power law dispersion relation (i.e., either pure gravity or pure capillary waves), which is not the case for the dispersion relation above (Equation \ref{RD}). For intermediate scales (i.e., $0.6 \lesssim \lambda \lesssim 5$ cm or  $6 \lesssim f \lesssim 40$ Hz for water waves) easily observed with tabletop experiments, both capillary and gravity forces are important and should be taken into account. This coexistence leads to several phenomena related to the non-monotonic feature of the phase velocity $\omega/k$ of linear waves such as Sommerfeld precursors \citep{FalconPRL2003}, parasitic capillarities \citep{FedorovPOF1998}, or Wilton waves \citep{HendersonJFM1987}. The minimum of the phase velocity corresponds to the transition between the gravity and capillary regimes and it occurs at $k_{gc}\equiv \sqrt{\rho g/\gamma}$ (the inverse of the capillary length) or, equivalently, at the frequency $f_{gc}\equiv(\rho/\gamma)^{1/4}{g}^{3/4}/(\sqrt{2}\pi)$. As the ratio $\rho/\gamma$ is constant for usual fluids, working in high-gravity or low-gravity environments provides a way to tune $f_{gc}$ significantly and expand the observation ranges of gravity wave turbulence \citep{CazaubielPRL2019} or of capillary wave turbulence \citep{FalconEPL2009}.
\begin{marginnote}[]
\entry{Crossover}{the wavelength $\lambda_{gc}$ or frequency $f_{gc}$ at which the gravity term $gk$ is equal to the capillary term $\frac{\gamma}{\rho}k^3$ in Equation (\ref{RD})}
\entry{Gravity wave regime}{wavelengths much larger than $\lambda_{gc}$ for which the restoring force is gravity}
\entry{Capillary wave regime}{wavelengths much smaller than $\lambda_{gc}$ for which the restoring force is capillarity}
\entry{Parasitic capillaries}{generated near steep crests of gravity- capillary waves as the result of a phase velocity matching between a long gravity-capillary wave and a shorter capillary one}
\end{marginnote}

\begin{figure}[t]
\includegraphics[height=6.5cm]{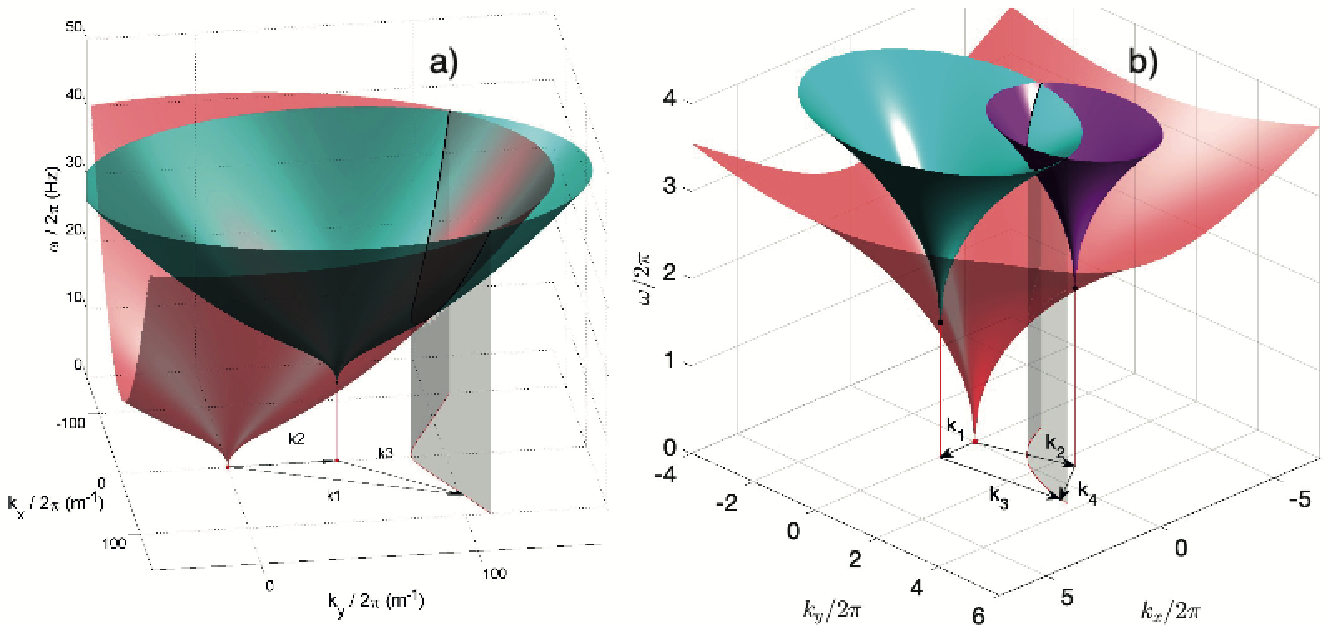}
\caption{Graphical solutions of Equation (\ref{resonance}) for $N$-wave resonant conditions. (\textit{a}) An $N=3$ solution of Equation (\ref{RD}) exists for gravity-capillary waves (shown) or for pure capillary waves ($\omega \sim k^{3/2}$; not shown), but (\textit{b}) only an $N=4$ solution exists for gravity waves ($\omega \sim k^{1/2}$). Panels adapted from (\textit{a}) \citet{AubourgPRF2016} and (\textit{b}) \citet{AubourgPRF2017} with permission from (\textit{a},\textit{b}) American Physical Society.}
\label{fig1}
\end{figure}

\section{NONLINEAR WAVE RESONANT INTERACTIONS}\label{interactions}
Resonant interactions between nonlinear waves constitute the fundamental mechanism that transfers energy in weak nonlinear wave turbulence. Generally speaking, $N$ waves interact with each other when the following conditions on angular frequencies $\omega_i$ and on wavevectors $\textbf{k}_i$ are satisfied
\begin{marginnote}[]
\entry{Wave steepness $\varepsilon$}{$\varepsilon \equiv k_p\eta_{\rm rms}$ with $k_p$ the wavenumber at the spectrum peak (maximum) and $\eta_{\rm rms}$ the standard deviation of the wave elevation} 
\end{marginnote}
\begin{equation}
\textbf{k}_1\pm \textbf{k}_2 \pm ...\pm \textbf{k}_N=0 \ \ {\rm and }\ \ \omega_1\pm \omega_2\pm ...\pm \omega_N=0{\rm ,} \ \ {\rm with }\ \  N\geq 3{\rm .}
\label{resonance}
\end{equation}
Each wave follows the dispersion relation (Equation \ref{RD}) as well, so that $\omega_i=\omega(|\mathbf k_i|)$. The magnitude of the nonlinear effects is quantified by the wave steepness $\varepsilon$, and the weak wave turbulence theory is based on an asymptotic expansion in $\varepsilon$. As a result, it only captures the dominant nonlinear interactions characterized by a single value of $N$: $N$ is thereafter the smallest integer for which Equation \ref{resonance} has non trivial solutions for a given $\omega(k)$ law. The different signs $\pm$ need to be the same in each instance of Equation \ref{resonance}.

As first suggested by \citet{Vedenov1967} and disseminated by \citet{NazarenkoBook2011}, Equation~\ref{resonance} can be solved graphically. Three-wave resonant interactions are only possible if $\omega(\textbf{k}_1=\textbf{k}_2+\textbf{k}_3)=\omega(\textbf{k}_2)+\omega(\textbf{k}_3)$ has a solution, i.e., if the surface $\omega(k_x,k_y)$ (in red in \textbf{Figure \ref{fig1}a}) has a nonempty intersection with the same surface (in blue) in a reference frame whose origin is on $\omega(\mathbf k_2)$. For pure power laws $\omega = a k^{b}$ this is only possible for $b>1$. In particular, three-wave resonant interactions occur for capillary waves ($b=3/2$), but they are forbidden for pure gravity waves ($b=1/2$) and thus four-wave interactions must be considered (see \textbf{Figure \ref{fig1}b}). However at scales near the crossover, nonlocal interactions can exist involving three-wave resonances between a gravity wave and two capillary waves due to the change of curvature of the dispersion relation (Equation \ref{RD}) \citep{McGoldrick1965,Simmons1969}. Furthermore, unidirectional interactions (i.e., collinear wave vectors $\mathbf k_i$) are allowed (which are not possible for pure power laws with $b\neq 1$).

\begin{marginnote}[]
\entry{Resonant interactions}{the two conditions of Equation \ref{resonance} are satisfied exactly with all waves following the dispersion relation}
\entry{Nonresonant interactions}{the two conditions are fulfilled but at least one of the involved Fourier modes is not a free wave (it does not follow the dispersion relation).}
\entry{Nonlocal interactions}{the wavelengths of interacting waves have very different orders of magnitude contrary to those involved in local interactions.}
\end{marginnote}

\begin{figure}[h]
\includegraphics[width=15cm]{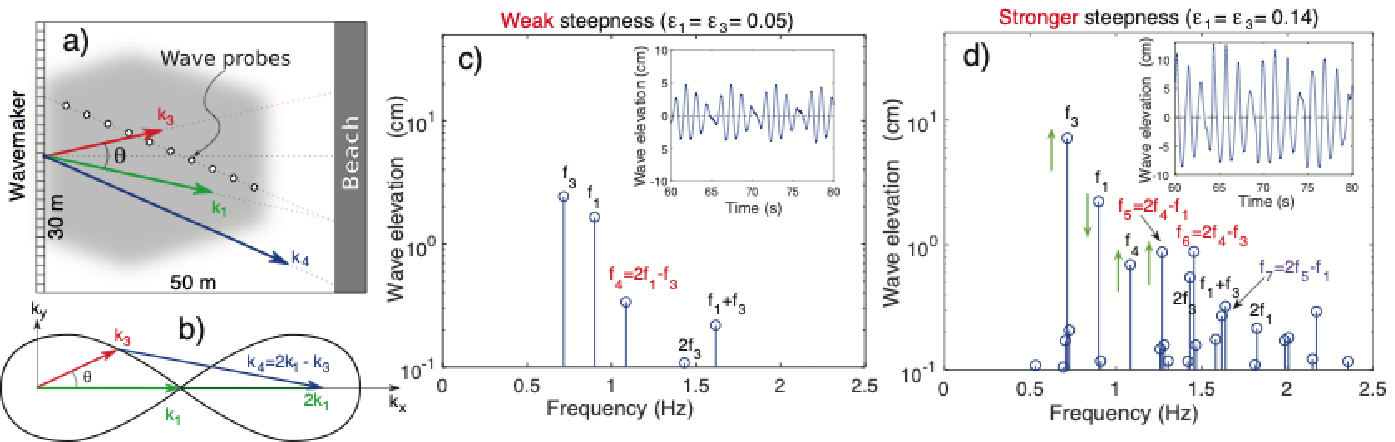} 
\caption{Nonlinear wave interaction. (\textit{a}) Mother gravity waves are generated at $\textbf{k}_1$, $\textbf{k}_2=\textbf{k}_1$ and $\textbf{k}_3$, crossing at an angle $\theta$ in a large-scale basin. Wave probes are located in the expected direction of the daughter wave ($\textbf{k}_4$) that should satisfy the four-wave resonant conditions ($2\textbf{k}_1-\textbf{k}_3=\textbf{k}_4$ and $2\omega_1-\omega_3=\omega_4$), graphically solved as (\textit{b}) \citet{PhillipsJFM1960} so-called figure of eight. (\textit{c}) For weak nonlinearity ($\varepsilon =0.05$), the wave elevation spectrum shows the observation of a daughter wave ($f_4=2f_1-f_3$) as well as second-order harmonics of the mother waves (2$f_3$, $f_1+f_3$, i.e., so-called bound waves). (\textit{d}) For stronger nonlinearity ($\varepsilon > 0.1$), there is pumping of a mother wave by daughter ones, and a cascade of quasi-resonances between mother waves ($f_1$ or $f_3$) and primary daughter waves ($f_4$), then with secondary daughter waves ($f_5$) and ($f_6$), and so on. Insets: Typical wave elevation monitored at the same distance.  Panels adapted from (\textit{a},\textit{b},\textit{c}) \citet{BonnefoyJFM2016} and (\textit{d}) \citet{BonnefoyHouille17} with permission from (\textit{a},\textit{b},\textit{c}) Cambridge University Press and (\textit{d}) EDP Sciences.}
\label{fig2}
\end{figure}



To extend early experiments on gravity wave resonances \citep{Longuet-Higgins1966,McGoldrick1966,Tomita89}, \citet{BonnefoyJFM2016} performed experiments on four-wave resonant interactions among surface gravity waves crossing 
in a large basin (see \textbf{Figure \ref{fig2}}a-b). This experimentally validated  
the theory of four-wave resonant interactions \citep{PhillipsJFM1960,Longuet-Higgins1962} for a wave steepness smaller than 0.1. Generating mother waves of a resonant quartet, they observed the growth of a daughter wave in the expected direction (see \textbf{Figure \ref{fig2}}c) and, notably, characterized its resonant properties (growth rate, response curve with the angle, and phase locking between waves). 
For stronger nonlinearities, departures from this weakly nonlinear theory were observed  such as additional daughter wave generation by nonresonant interactions ~\citep{BonnefoyHouille17}, which have been well described theoretically by \citet{Zakharov1968} (see \textbf{Figure \ref{fig2}}d). 


For capillary waves, the inviscid theory of three-wave resonant interactions \citep{McGoldrick1965,Simmons1969} has been tested qualitatively in early experiments \citep{McGoldrick1970,HendersonJFM1987} and then verified quantitatively experimentally and extended to more general cases by \citet{HaudinPRE2016} and \citet{AbellaPS2019}.

For pure gravity or capillary waves, as for all dispersive waves, resonant interactions following Equation \ref{resonance} involve waves propagating in distinct directions. However, close to the gravity-capillary crossover, unidirectional resonant interactions involving a gravity wave and two capillary waves are possible and have been observed experimentally to be the most active \citep{AubourgPRL2015,AubourgPRF2016}.

\begin{marginnote}[]
\entry{Quasi-resonant interactions}{the two conditions of Equation \ref{resonance} are approximately fulfilled, with weak nonlinear or dissipative corrections to the dispersion relation}
\end{marginnote}

Although weak turbulence theory is restricted to strict resonant wave interactions in the limit of vanishing $\varepsilon$, quasi-resonant coupling among waves is also found to play a significant role in experiments. As discussed in Section~\ref{2Dspectrum}, nonlinear widening of the dispersion relation at a nonzero value of $\varepsilon$ enables approximate resonances. Another physical mechanism is dissipation that increases the resonance bandwidth (as for the damped forced oscillator) and authorizes three-wave interactions at nonresonant angles \citep{CazaubielPRF2019}. 




\section{WEAK TURBULENCE THEORY}\label{WWT}
\subsection{Kinetic Equation}
Details on the development of weak turbulence theory can be found for instance in the textbooks by \citet{ZakharovBook1992} and  \citet{NazarenkoBook2011} and in the review by \citet{Newell2011}. The Hamiltonian equation in Fourier space reads $i\frac{\partial a_\textbf{k}}{\partial t} = \frac{\partial H}{\partial a_\textbf{k}^*}$, with $H$ the Hamiltonian of the system and $a_\textbf{k}$ the canonical variables associated with complex wave amplitudes in Fourier space. An asymptotic expansion of the Hamiltonian, using a scale separation hypothesis between the slow time of nonlinear interactions and the fast time of linear wave oscillations, leads to
\begin{equation}
\begin{gathered}
i\frac{\partial a_\textbf{k}}{\partial t} = \frac{\partial H}{\partial a_\textbf{k}^*}
=\omega a_\textbf{k} + \varepsilon \int V_{k,k_1,k_2} a_{\textbf{k}_1} a_{\textbf{k}_2} \delta (\textbf{k}_1+\textbf{k}_2-\textbf{k})d\textbf{k}_1 \textbf{k}_2\\
+\varepsilon^2\int W_{k,k_1,k_2,k_3} a_{\textbf{k}_1}a_{\textbf{k}_2} a_{\textbf{k}_3}\delta (\textbf{k}_1+\textbf{k}_2+\textbf{k}_3-\textbf{k})d\textbf{k}_1 \textbf{k}_2 \textbf{k}_3 + ...,
\end{gathered}
\label{Hamiltonien}
\end{equation}
with $V_{k,k_1,k_2}$ the three-wave interaction coefficient and $W_{k,k_1,k_2,k_3}$ the four-wave interaction coefficient  \citep{Hasselmann62,ZakharovBook1992,NazarenkoBook2011}. For $\varepsilon \ll 1$, one can consider only the smallest nonzero coefficient in this development. As discussed above, $N=3$ for capillary waves and $N=4$ for gravity waves. To reach statistical properties, weak turbulence theory computes the second-order moment of the canonical variable $\langle{a_{\textbf{k}}a_\textbf{k'}}\rangle$ using the random phase hypothesis (wave phase and amplitude are assumed quasi-Gaussian) \citep{NazarenkoBook2011} or the hierarchy of the cumulants of the canonical variables \citep{Newell2001}, where $\langle \cdot \rangle$ denotes a statistical average. Assuming spatial homogeneity 
and based on the linear and nonlinear timescale separation hypothesis, an asymptotic closure arises in the limit of infinite system size and of vanishing nonlinearity. The resulting kinetic equation describing the long-time evolution of the wave action spectrum $n_\textbf{k}=\langle a_\textbf{k} a_{\textbf{k}}^*\rangle$ reads, for three-wave interactions (capillary case)
\begin{equation}
\begin{aligned}
\frac{\partial n_\textbf{k}}{\partial t} = 4\pi \varepsilon^2 \int |V_{k,k_
1,k_2}|^2  n_{\textbf{k}} n_{\textbf{k}_1} n_{{\textbf{k}_2}}\delta (\textbf{k}-\textbf{k}_1-\textbf{k}_2)
\left[\left(\frac{1}{n_{{\textbf{k}}}}-\frac{1}{n_{{\textbf{k}_1}}}-\frac{1}{n_{{\textbf{k}_2}}}\right)\delta (\omega -\omega_1-\omega_2)+\right.\\
 \left(\frac{1}{n_{{\textbf{k}}}}-\frac{1}{n_{{\textbf{k}_1}}}+\frac{1}{n_{{\textbf{k}_2}}}\right)\delta (\omega_1 -\omega-\omega_2)+
 \left.\left(\frac{1}{n_{{\textbf{k}}}}+\frac{1}{n_{{\textbf{k}_1}}}-\frac{1}{n_{{\textbf{k}_2}}}\right)\delta (\omega_2 -\omega_1-\omega)\right]d\textbf{k}_1 d\textbf{k}_2\ {\rm ,}
\end{aligned}
\label{cinetique3}
\end{equation}
\begin{marginnote}[]
\entry{Kinetic equation}{equation for the slow temporal evolution of the wave action spectrum $n_{\mathbf k}$}
\end{marginnote}
and, for four-wave interactions (gravity case),
\begin{equation}
\begin{aligned}
\frac{\partial n_\textbf{k}}{\partial t} =4\pi \varepsilon^4 \int|W_{k,k_1,k_2,k_3}|^2 n_\textbf{k} n_{\textbf{k}_1} n_{\textbf{k}_2} n_{\textbf{k}_3}\delta (\textbf{k}+\textbf{k}_1-\textbf{k}_2-\textbf{k}_3)\left[\frac{1}{n_{\textbf{k}}}+\frac{1}{n_{\textbf{k}_1}}-\frac{1}{n_{\textbf{k}_2}}-\frac{1}{n_{\textbf{k}_3}}\right] \\
\delta(\omega+\omega_1-\omega_2-\omega_3)d\textbf{k}_1d\textbf{k}_2 d\textbf{k}_3\ {\rm ,}
\end{aligned}
\label{cinetique4}
\end{equation}
Note that the collision integral contributes to the spectrum evolution only when the resonant interaction conditions are satisfied due to Dirac's $\delta$ functions. For the complete gravity-capillary system, most likely both terms should be taken into account, although it has never been investigated.

\subsection{Constant Flux Solutions} \label{Solution}
By definition of canonical variables, the spectral energy density (or wave energy spectrum) $E_k$ is related to the wave action spectrum $n_k$ by $E_k=\omega(k)n_k$, the total wave energy $E = \int E_k dk$ being conserved. The energy flux $P$ is defined by the following balance: $\frac{\partial E_k}{\partial t} +\frac{\partial P}{\partial k} = 0$. Stationary solutions of the kinetic equation cancel the collision integral and thus correspond to a constant energy flux $P$ across scales (in practice, between the energy source and sink). For power law dispersion relations, $\omega = ak^b$, Zakharov's transformation \citep{ZakharovBook1992,NazarenkoBook2011} provides the stationary isotropic solutions as power laws in $k$
\begin{equation}
n_k = 2\pi C_0 P ^{1/(N-1)}a^{-\alpha} k^{-\beta},
\label{KZn}
\end{equation}
where $N$ is the leading-order interaction in the system, $C_0$, $\alpha$, and $\beta$ are constants that can be computed and that depend on the wave system considered. By analogy to the Kolmogorov spectrum in 3D hydrodynamic turbulence, these solutions are called the Kolmogorov-Zakharov (KZ) spectra.

Since the exact analytical computation of the above solutions is rather long and technical, one way to find the KZ spectrum scalings is to use dimensional analysis \citep{ZakharovBook1992,Connaughton2003,NazarenkoBook2011}. Let us consider waves propagating in two dimensions according to $\omega = ak^b$, where the dimension of $a$ is $\left [L^b T^{-1}\right]$. The dimension of the energy density $E_k$, normalized by unit of surface and density, is $\left[L^4 T^{-2}\right]$. The energy flux $P$, similarly normalized, has the dimension $\left[L^3 T^{-3}\right]$.
For a system dominated by $N$ wave interactions, the energy flux is proportional to the power $N-1$ of the spectral energy density (and thus $E_k\sim P^{\frac{1}{N-1}}$) \citep{KraichnanPOF1965,Connaughton2003}. Dimensional analysis thus yields
\begin{marginnote}[]
\entry{Kolmogorov-Zakharov (KZ) spectra}{ \\ constant flux, out-of-equilibrium, stationary solutions of the kinetic equation}
\end{marginnote}
\begin{equation}
E_k \sim P ^{\frac{1}{N-1}}a^\xi k^\zeta {\rm , \ \ with}\ \  \ \xi = 2 -3/(N-1)\ \ {\rm and}\ \ \ \zeta = 2b -4+(3-3b)/(N-1) {\rm .}
\label{direct}
\end{equation}
One has also $\beta=b-\zeta$ and $\alpha=1-\xi$. Most experiments rather consider the power spectral density $S_k=2\pi k\langle \left|\iint_0^L\eta(x,y)e^{i(k_xx+k_yy)}dxdy \right|^2\rangle /L^2$ or $S_\omega=\langle\left|\int_0^T\eta(t)e^{i(\omega t)}dt \right|^2\rangle /T$ of the measured wave elevation $\eta(x,y)$ or $\eta(t)$, $L$ being the window size and $T$ the recording time.
$S_k$ is related to the energy spectrum by $E_k^g=\frac{1}{2} g S_k^g$ for gravity waves, and $E_k^c=\frac{\gamma}{2\rho}k^2 S_k^c$ for capillary waves. These densities can be changed to frequency space using $E_k dk =E_\omega d\omega$ and $S_k dk =S_\omega d\omega$ and the linear dispersion relation.
For deep-water gravity waves ($N=4$, $b=1/2$, $a = \sqrt{g}$), Equation \ref{direct} thus yields the spectrum predictions of the direct energy cascade
\begin{equation}
E_k^g\sim P^{1/3} g^{1/2}k^{-5/2},\ \ \ \ S_k^g\sim P^{1/3}g^{-1/2}k^{-5/2}, \ \ \ \ S_\omega^g\sim P^{1/3} g \omega^{-4}.
\label{directg}
\end{equation}
The exact solution was derived by \citet{Zakharov1967grav}. For capillary waves [$N=3$, $b = 3/2$, $a =(\gamma/\rho)^{1/2}$], the spectrum predictions are

\begin{equation}
E_k^c\sim P^{1/2}\left( \frac{\gamma}{\rho}\right)^{1/4}k^{-7/4}{\rm ,} \ \  \ S_k^c\sim P^{1/2}\left(\frac{\gamma}{\rho}\right)^{-3/4}k^{-15/4}{\rm ,}\ \ \ S_\omega^c\sim P^{1/2} \left(\frac{\gamma}{\rho}\right)^{1/6} \omega^{-17/6}{\rm .}
\label{SpecCap}
\end{equation}
The exact solution was derived by \citet{Zakharov1967}. 

The nondimensional KZ constant $C_0$ was estimated experimentally for gravity waves by \citet{DeikeJFM2015} and found to be of the same order of magnitude as the theoretical value ($C^g_0=2.75$) estimated by \citet{Zakharov2010}. For capillary waves, the  KZ constant was first analytically evaluated as 9.85 by \citet{Pushkarev2000} and corrected by \citet{PanYueJFM2017} to $C^c_0=6.97$. Using a low dissipation level, direct numerical simulation by \citet{DeikePRL2014} led to $C^c_0=5\pm 1$, whereas \citet{PanPRL2014} found a value that depends on the system's finite size. Experimental estimation of the KZ capillary constant (see Section~\ref{capillary}) led to $C^c_0 \approx 0.5$ \citep{DeikePRE2014}.

For the full gravity-capillary system, since the dispersion relation is not a pure power law, so far no analytical solution for the KZ spectrum exists, and dimensional analysis is not conclusive. One may expect to recover the pure gravity or capillary cases at scales far from the crossover but the connection between the two solutions in the intermediate region remains unclear. Because the scalings in $P$ of the two pure cases are different, one may expect the transition between both regimes to occur at distinct scales when changing the energy flux. When equating the two KZ spectra of Equations \ref{directg} and \ref{SpecCap}, one obtains the transition between the two spectra $S_k$ at $k=k_{gc}(C_0^c/C_0^g)^{4/5} (P/P_b)^{2/15}$, where $P_b=(\gamma g/\rho)^{3/4}$ is the energy flux breaking weak turbulence (see Section~\ref{taunl}). The transition should thus slightly increase with $P$ and be equal to $k_{gc}$ only for $P=P_b(C_0^g/C_0^c)^6\simeq P_b/265$. 

%

When $N$ is even and a conservation of the number of interacting waves occurs (as for gravity waves), the total wave action $\mathcal{N} = \int n_k dk$ is conserved as well, and the wave action flux $Q$ is defined as $\frac{\partial n_k}{\partial t}+\frac{\partial Q}{\partial k}=0$. An inverse cascade (from small scales to large ones) is then predicted characterized by a constant wave action flux through the scales once a stationary state is reached. Since $[Q]=[P]/[\omega]$, the dimension of $Q$ is $[L^3 T^{-2}]$, and dimensional analysis leads to the inverse cascade spectrum
\begin{marginnote}[]
\entry{Conservation of the number of interacting waves}{for $N=3$, the number of interacting waves is never conserved ($2\leftrightarrow1$ process); for $N=4$, it is either conserved ($2 \leftrightarrow 2$) or not ($3 \leftrightarrow 1$)}
\end{marginnote}
\begin{equation}
E^i_k \sim Q^{\frac{1}{N-1}} a^\xi k^\zeta {\rm , \ \ with}\ \  \ \xi =2-2/(N-1)\ \ {\rm and}\ \ \ \zeta =2b-4 +(3-2b)/(N-1) {\rm .}
\end{equation}
For deep-water gravity waves ($N=4$, $b=1/2$, $a = \sqrt{g}$), the inverse cascade spectra are
\begin{equation}\label{inverse}
E^i_k \sim Q^{1/3} g^{2/3} k^{-7/3}{\rm ,}\ \ \ \ S_k^i \sim Q^{1/3} g^{-1/3} k^{-7/3}{\rm ,} \ \ \ \ S_\omega^i\sim Q^{1/3} g \omega^{-11/3}{\rm .}
\end{equation}
The exact solution was derived by \citet{Zakharov82GravInv}.

\subsection{Zero-Flux or Independent-Flux Solutions}\label{zerofluxsolution}
Other solutions of the wave action spectra exist beyond those of Section~\ref{Solution}. For example, for capillary waves, no inverse cascade is predicted ($N=3$ is odd), and the dynamics at scales larger than the forcing scale is thus expected to follow the statistical (or ``thermodynamic'') equilibrium state, that is, the kinetic energy equipartition among the Fourier modes with no wave action flux towards large scales \citep{BalkovskyPRE1995}. The spectrum of large-scale capillary wave turbulence is thus predicted as $S^{th}_k=k_BT/(2\pi \sigma k)$ or $S^{th}_{\omega}=2k_BT/(3\sigma \omega)$ \citep{MichelPRL2017} with $k_B$ the Boltzmann constant and $T$ a constant effective temperature related to the total energy within this out-of-equilibrium stationary state. This is the analog of the Rayleigh-Jeans spectrum of the blackbody radiation.
 
Beyond weak turbulence, a flux-independent solution 
for gravity waves was proposed dimensionally by \citet{PhillipsJFM1958} as $S^{Ph}_{\omega} \sim P^0g^2\omega^{-5}$. It is interpreted as a saturated spectrum, a situation that in practice corresponds to nonlocal interactions where localized coherent structures (whitecaps, wave breaking) associated with steep gravity waves dissipate all the injected energy \citep{Newell2011}. At intermediate stages (between weak turbulence and saturation) \citet{Kuznetsov04} proposed that the spectrum would be proportional to the density $n$ of singularities with an exponent in $k$ that depends on the geometry of the structures. If these dissipative structures occur along lines rather than locally (sharp-crested waves) then $S^K_{\omega} \sim n\omega^{-4}$ is expected \citep{Kuznetsov04,NazarenkoJFM2010}. Numerical simulations have shown some evidence of Phillips' spectrum in the inverse cascade regime \citep{KorotkevitchPRL2008,KorotkevitchMCS2008}. 

\section{TIMESCALES OF WAVE TURBULENCE}\label{Timescale}
Weak turbulence theory assumes a timescale separation $\tau_{lin}(k)\ll \tau_{nl}(k) \ll  [\tau_{diss}(k)$ and $\tau_{disc}(k)]$. In the whole inertial range, the timescale of nonlinear interactions between waves, $\tau_{nl}$, is assumed to be large compared with the linear time, $\tau_{lin}=1/\omega$, so that the nonlinear evolution is slow compared with the fast linear oscillations of the waves. In addition, $\tau_{nl}$ must be short compared to the typical dissipation time $\tau_{diss}$ and the discreteness time $\tau_{disc}$. Let us discuss all these time scales.

\subsection{Nonlinear Timescale} \label{taunl}
From scaling arguments on the kinetic equation, the nonlinear interaction time reads \citep{Newell2011}
\begin{equation}
\tau^g_{nl}\sim P^{-2/3}{{g}^{1/2}}k^{-3/2} {\rm \ \ (gravity), \ \ and}\ \   \tau^{c}_{nl}\sim P^{-1/2}(\gamma/\rho)^{1/4}k^{-3/4}{\rm \ \ (capillary).} \label{eq:taunl}
\end{equation}
For gravity waves, one must have $\tau_{lin}/\tau^g_{nl} \sim P^{2/3}k^{1}/g \ll 1$ \citep{Newell2011}. As this ratio increases with $k$, breakdown of the weak nonlinearity hypothesis is expected to occur at {\em small scales}  for $k>k^g_b \sim g/P^{2/3}$. By contrast, for capillary waves, breakdown occurs at {\em large scales} since the ratio $\tau_{lin}/\tau^c_{nl}\sim P^{1/2}(k\gamma/\rho)^{-3/4} \ll 1$ decreases with $k$ and thus exceeds one for $k<k^c_b \sim P^{2/3}\rho/\gamma$. 
At small enough $P$ one has $k^g_b>k^c_b$, and thus a weak regime of gravity-capillary turbulence may develop at all scales. Equating these two breaking scales, $k_b=k^g_b=k^c_b$, leads to a critical energy flux, $P_{b}=(\gamma g/\rho)^{3/4}$, that breaks weak gravity-capillary wave turbulence \citep{NewellPRL92}. For $P>P_b$, one has $k^g_b<k^c_b$ and a window in $k$-space exists (typically $k\in [k^g_b,k^c_b]$, near the gravity-capillary transition) where the dynamics is expected to be dominated by strongly nonlinear structures (white caps, sharp-crested waves) \citep{NewellPRL92,Connaughton2003} or by nonlocal interactions, such as parasitic capillary wave generation \citep{FedorovPOF1998}, as evidenced experimentally by the occurence of stochastic energy bursts transferring wave energy non-locally from gravity waves to all capillary spatial scales quasi-instantaneously \citep{Berhanu2013,BerhanuJFM2018}. However, no such transition from a weak turbulence spectrum to a strong turbulence spectrum (Phillips' spectrum) at high wave numbers has been reported experimentally so far. This independent-flux solution (i.e., Phillips' spectrum of sharp, crested waves) has a $k$-independent ratio, $\tau_{lin}/\tau_{nl} \sim k^0$ \citep{Newell2011}. For usual fluids, $P_b$ is roughly constant, about $4 200$ cm$^3$/s$^3$. Experimentally, the cascading energy flux $P$ can be indirectly measured (see Section~\ref{capillary}) and is found to be more than one order of magnitude smaller than the critical flux $P_b$ \citep{DeikeJFM2015,CazaubielPRL2019}. The value of $\tau_{nl}$ can be measured with a local probe by decaying wave turbulence experiments either in the gravity regime \citep{BedardJETP2013,DeikeJFM2015} or in the gravity-capillary regime \citep{CazaubielPRL2019}, and by the broadening of the dispersion relation in stationary experiments using space-time measurements \citep{HerbertPRL2010,BerhanuJFM2018}. In the latter case, the width of the energy concentration around the dispersion relation can be quantified either in frequency space by $\delta \omega\propto 1/\tau_{nl}$ or in wavenumber space by $\delta k\propto \frac{\partial k}{\partial \omega}\delta \omega$.

Similarly, for the inverse cascade of gravity waves, the nonlinear timescale reads $\tau^{i}_{nl}\sim  g^{1/6}Q^{-2/3}k^{-11/6}$, and thus we have the ratio $\tau_{lin}/\tau^i_{nl} \sim Q^{2/3}g^{-2/3}k^{4/3} \ll 1$ \citep{Newell2011}. As this ratio increases with $k$ or with the wave action flux $Q$, breakdown of weak turbulence is expected to occur at small scales when $k> k_b= \sqrt{g/Q}$ or for $Q > Q_{b}=g/k^{2}$.  Here also, wave action flux in experiments is much smaller than this critical value \citep{FalconPRL2020}.

\subsection{Dissipation Time}
The scale separation $\tau_{nl}(k) \ll  \tau_{diss}(k)$ is taken for granted in the theory but it is not so straightforward in real life. Energy dissipation in water waves occurs mainly through three distinct mechanisms:  viscous linear damping (very small for large-scale waves $\lambda \gtrsim 0.5$ m), energy extraction by generation of parasitic capillaries near steep crests of longer waves \citep{LonguetHiggins1963,FedorovPOF1998}, and wavebreaking (i.e., a multivalued interface) at very large steepnesses. When assuming a stress-free water/air interface, the typical linear viscous dissipation timescale is due the water bulk viscosity and thus we have $\tau_{diss}=1/(2\nu k^2)$  \citep{Lamb1932,Miles1967,DeikePRE2012}. For a contaminated interface, the air/water surface boundary layer due to an inextensible film gives $\tau^s_{diss}=2\sqrt{2}/(k\sqrt{\nu \omega})$ \citep{VanDorn,HendersonJFM1990} [which is negligible for $f \lesssim 2$ Hz \citep{CampagnePRF2018}]. The boundary layer on the lateral walls (for experiments) yields $\tau^L_{diss}=2\sqrt{2}L_xL_y/[3\sqrt{\nu\omega}(L_x+L_y)]$ \citep{Miles1967,CazaubielPRL2019}, whereas the bottom boundary layer dissipation is negligible for deep-water waves. 
The term $\tau_{diss}$ is usually measured with a local probe by decaying wave turbulence experiments either in the gravity regime \citep{BedardJETP2013,DeikeJFM2015}, in the gravity-capillary regime \citep{CazaubielPRL2019}, or in the capillary regime \citep{DeikePRE2012}. 

\subsection{Discreteness Time}\label{discret}
Finite-size effects often play a role in wave turbulence experiments, as the presence of confining lateral walls cannot be avoided. A closed basin exhibits eigenmodes that depend on its size and geometry. Indeed, the boundary conditions lead to a discretization of possible wave vectors. For example, for a rectangular basin of size $L_x$ and $L_y$, the eigenmodes are $k_d=\sqrt{\left(m\pi/L_x\right)^2+\left(n\pi/L_y\right)^2}$ with $m,n \in \mathbb{N}$  \citep{Lamb1932}. 
The discreteness time $\tau_{disc}$ can be computed as the number of eigenmodes found in a frequency band divided by this bandwidth \citep{FalconPRL2020}. When the nonlinear spectral widening $\delta k$ is greater than the half separation  $\Delta k /2$ between adjacent eigenmodes, they are no longer separated and this prevents any effect of discreteness. It occurs when $\tau_{nl}(k) < 2\tau_{disc}(k)$, with $\tau_{disc}=1/\Delta \omega$ and $\Delta \omega=(\partial \omega / \partial k) \Delta k$. In this case, one expects to recover a kinetic regime with effectively continuously varying wavenumbers as in the limit of infinite system size considered in the theory. In the opposite case, $\tau_{nl}(k) > 2\tau_{disc}(k)$, discrete wave turbulence is expected. The intermediate regime, $\tau_{nl} \sim 2\tau_{disc}$, is called ``frozen'' or ``mesoscopic'' wave turbulence \citep{NazarenkoBook2011}. For instance, in a system of size $L$, one has $\Delta k=\pi/L$, and the discreteness time reads $\tau^g_{disc}=(2L/\pi) \sqrt{k/g}$ for gravity waves and $\tau^c_{disc}=[2L/(3\pi)]\sqrt{\rho/(\gamma k)}$ for capillary waves \citep{Ricard2021b}. The frozen scale occurs for $\tau_{nl}(k_{fr}) =2\tau_{disc}(k_{fr})$, that is, using both parts of Equation \ref{eq:taunl}, $k^g_{fr}=\sqrt{\pi g/(4L)}P^{-1/3}$ and $k^c_{fr}=[3\pi/(2L)]^4(\gamma/\rho)^3P^{-2}$ for each regime \citep{Ricard2021b}. The finite-system size effects are thus more significant when L or P decreases. Taking these effects into account in theories is an important challenge \citep{Zakharov2005,Nazarenko06,KartashovaPRE08,LvovPRE10,PanYueJFM2017,HrabskiPRE2020}.



\subsection{Timescale Separation}
The physical properties of water provide constraints on the scale separation. For large-scale gravity waves ($\lambda > 1$ m), one has $\tau_{lin}/\tau_{diss} < 10^{-5}$ since the dissipation due to bulk viscosity is almost negligible \citep{CampagnePRF2018}, and one thus expects a proper scale separation in field observations at these scales. In laboratory experiments, the system size most often restricts the studied scales to wavelengths below 1 m, even in large-scale tanks. Near the crossover and for smaller wavelengths, the ratio is $\tau_{lin}/\tau_{diss} \gtrsim 10^{-2}$ for water, and further increases in the case of surface contamination \citep{CampagnePRF2018}. This means that the forcing (and so the wave steepness) must be high enough to reach an adequate scale separation ($\tau_{lin} \ll \tau_{nl} \ll \tau_{diss}$) at these wavelengths, at the risk of not being so weakly nonlinear (typically $\varepsilon\simeq 0.05 - 0.1$). 
To decrease capillary viscous dissipation, researchers have performed experiments with mercury  \citep{FalconPRL2007a,FalconPRL2007b,FalconPRL2008,Ricard2021}, or with liquid hydrogen \citep{BrazhnikovEPL2002,Kolmakov2009}.
A direct estimation of the nonlinear timescale is not straightforward and was accomplished only in a few cases (see Section~\ref{taunl}). The timescale separation was found to be well validated experimentally for gravity wave turbulence \citep{DeikeJFM2015,FalconPRL2020} and for gravity-capillary wave turbulence \citep{CazaubielPRL2019}, as well as numerically for pure capillary wave turbulence \citep{DeikePRL2014} [see also \citet{DeikeJFM2013} and \citet{Miquel2014} for such tests in other experimental wave turbulence systems]. However, when the finite-size effects are significant, the nonlinear and dissipative timescales are found to be independent of the scale, contrary to weak turbulence predictions \citep{CazaubielPRL2019}.    



\begin{figure}[t]
\includegraphics[width=16cm]{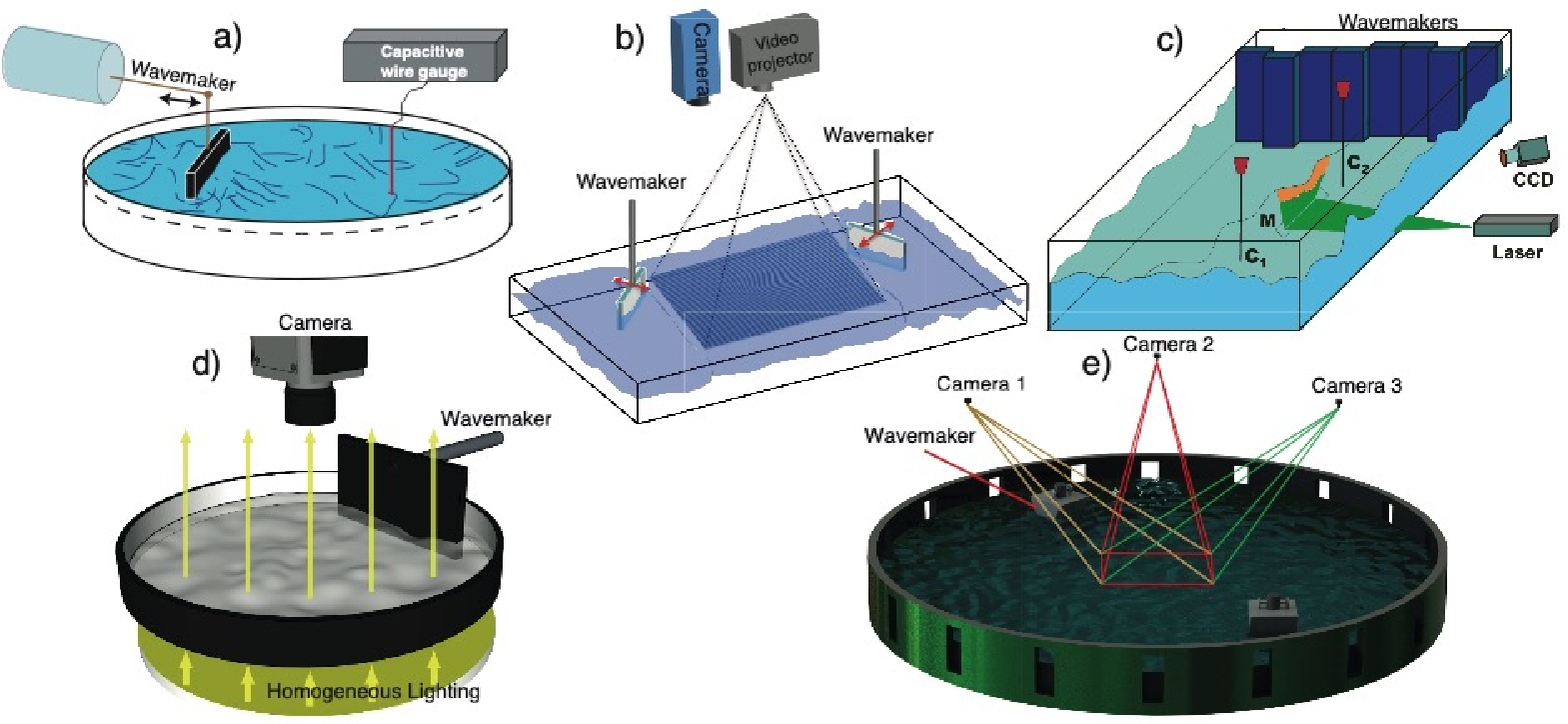} 
\caption{Typical surface wave elevation measurements: (\textit{a}) capacitive wire gauge, (\textit{b}) 3D wave field reconstruction by Fourier Transform Profilometry (FTP), (\textit{c}) 2D spatial profile, (\textit{d}) 3D Diffusing Light Profilometry (DLP), and (\textit{e}) 3D stereo-PIV. Panels adapted from (\textit{a}) \citet{CazaubielPRL2019}, (\textit{b}) courtesy from P. Cobelli, (\textit{c}) \citet{NazarenkoJFM2010}, (\textit{d}) courtesy from J.-B. Gorce, (\textit{e}) \citet{AubourgPRF2017} with permission from (\textit{a},\textit{d},\textit{e}) American Physical Society.} 
\label{fig3}
\end{figure} 

\section{EXPERIMENTAL METHODS}\label{methods}
Water waves are commonly generated by a localized forcing using a wave maker made of one or multiple independently controlled paddles (see \textbf{Figure \ref{fig3}}). Injected power into the fluid can be measured \citep{FalconPRL2008}, as can the energy flux $P$, indirectly (see Section~\ref{capillary}).

\subsection{Single-Point Measurements}In field observations, surface wave elevations are usually measured by buoys, lidar or microwave radars. In laboratory experiments, resistive or capacitive wire gauges are widely used (see \textbf{Figure \ref{fig3}a}). Capacitive probes are made of a thin insulated wire in water, considered as an annular capacitor with a capacity proportional to the immersed length of the wire. Although intrusive, they are easy to implement and have a wide measurement range (from 10 $\mu$m to tens of cm with a frequency cutoff up to a few hundred Hz and no limitation in wave steepness) \citep{FalconPRL2007a,FalconPRL2007b,DeikePRE2012}. They are more adequate for small-scale resolution than resistive probes, which are restricted to gravity wave studies. The resistive probe accuracy for the height detection is $\lesssim 100\ \mu$m with a rather low frequency cutoff $\lesssim 20$ Hz \citep{CazaubielPRL2019}. 

To avoid possible disruption of the wave field by gauges, several authors have used nonintrusive optical methods based on tracking by a position-sensitive detector of the partial adsorption \citep{Henry2000}, reflection \citep{Lommer02,BrazhnikovEPL2002,Kolmakov2009}, or refraction \citep{Snouck2009} of a laser beam at one point of the fluid surface to study capillary wave turbulence with a parametric forcing. Other authors have used single-point laser Doppler vibrometers to study capillary wave turbulence \citep{Holt96}, depth-induced properties in gravity-capillary wave turbulence \citep{FalconEPL11}, wave turbulence in a two-layer fluid \citep{IssenmannEPL2016}, and gravity-capillary wave resonant interactions \citep{HaudinPRE2016,CazaubielPRF2019}. 
Laser vibrometry consists in a reference laser beam that interferes with light backscattered by the moving free surface. It infers the normal wave velocity by the Doppler effect and the wave elevation by interferometry, with high displacement measurement accuracy (up to $\sim 0.3\ \mu$m or $0.1\ \mu$m/s), thin spatial extension of the probe region of the order of $10 \mu$m, that is a few times the laser beam diameter, and a high temporal dynamics (timescales down to $\sim  \mu$s). It requires the addition of light scattering particles in water and is limited to low wave steepness ($< 0.1$). 

Beyond the above Eulerian specifications of the wave field, recent articles report the use of particle tracking velocimetry of Lagrangian buoyant particles or floaters to study gravity-capillary wave turbulence \citep{DelGrossoPRF2019,Cabrera2021}.

\subsection{Space-Time Measurements} \label{spacetime}
Simultaneous measurements in the time and space domains enable one to discriminate weakly nonlinear waves that verify the dispersion relation from other more complex dynamics.

Space-and-time-resolved imaging of the free surface along a line can be achieved by using a laser sheet impinging the water surface  \citep{LukaschukPRL09} (see \textbf{Figure \ref{fig3}c}) or scanning a laser beam refracted by the free surface \citep{Snouck2009}. In linear flumes with transparent side walls, laterally positioned cameras can image the 2D spatial profile along the flume \citep{Redor2020,Ricard2021}. 

Nowadays, full 3D spatial wavefield reconstructions are achieved using high-speed cameras.
The Fourier Transform Profilometry (FTP), introduced by \citet{TakedaAO1983} and further developed by \citet{CobelliEF2009}, was used by \citet{HerbertPRL2010} to obtain the nonlinear dispersion relation and spatial statistics of gravity-capillary wave turbulence. It is now the most widespread 3D reconstruction technique for small-scale experiments \citep{CobelliPRL2009,CobelliPRL2011,DeikeJFM2013,AubourgPRL2015}. The principle is to project a grayscale pattern made of parallel lines at the water surface. A fast camera records the pattern deformed by the waves; the water height can be recovered by a demodulation algorithm (see \textbf{Figure \ref{fig3}b}). The spatial horizontal resolution is of the order of the distance between the projected lines (few mm, typically) and the resolution of the measurement of the water elevation is about $100\ \mu$m. A white dye must be added to render the water diffusive to light. A paint dye was first used \citep{HerbertPRL2010}, followed by titanium dioxide (TiO$_2$) particles, which led to much lower modifications of the fluid properties (surface contamination, viscosity) \citep{Przadka2012}. 

Diffusing light photography or profilometry (DLP) is a technique more adapted to capillary wave turbulence since its horizontal and vertical resolution are higher ($\sim 10\ \mu$m) than those of FTP. \citet{Wright1996,Wright1997} introduced DLP to study capillary wave turbulence, but the spatial wave height reconstruction was achieved using photographs with no temporal resolution (only collections of snapshots). By associating this technique with a fast camera, \citet{Berhanu2013}, \citet{HaudinPRE2016}, and \citet{CazaubielPRF2019} obtained the full space-and-time-resolved measurements of gravity-capillary wave turbulence (see \textbf{Figure \ref{fig3}d}). This optical method is based on the light absorption of a diffusing fluid (water and microspheres). The surface topography is reconstructed from the variations of the light intensity transmitted through the liquid illuminated from below and captured by a fast camera from above. 
Contrary to usual optical methods based on the wave slope measurement (reflection or refraction), DLP works well for steeply sloping waves and is thus well adapted to study strong capillary wave turbulence \citep{BerhanuJFM2018}. A similar absorption technique was implemented at an interface between two index-matched liquids (a transparent upper liquid and a dyed lower liquid) to reconstruct Faraday surface wave patterns \citep{Kityk2004}.

Synthetic Schlieren was first developed to image internal waves \citep{Peters1985,Dalziel2000}, and has since been applied to image water waves \citep{Kurata1990,Moisy2009} and, more recently, to test surface three-wave resonant interactions \citep{AbellaPS2019}. This method is based on the analysis of the image of a random dot pattern (placed below the wave tank) refracted trough the water surface. The reconstruction of the wave field is obtained using a digital cross-correlation PIV (particle image velocimetry)-type algorithm. Despite its extreme sensitivity ($\sim 1-10\ \mu$m), this method based on light refraction provides the gradient of wave height and is thus limited to small wave steepness (to prevent the formation of caustics) and small wave amplitudes.

Time-and-space-resolved wave field measurements from video using multiple camera views are currently booming, notably for gravity waves, as in stereoscopy \citep{Benetazzo2006,Leckler15,ZavadskyPOF2017} or stereo-PIV \citep{Prasad2000,Turney2009,AubourgPRF2017} (the latter requiring tracers, with a vertical resolution $\sim $ mm, and a horizontal resolution of a few cm; see \textbf{Figure \ref{fig3}e}). 

\begin{figure}[h]
\includegraphics[width=16cm]{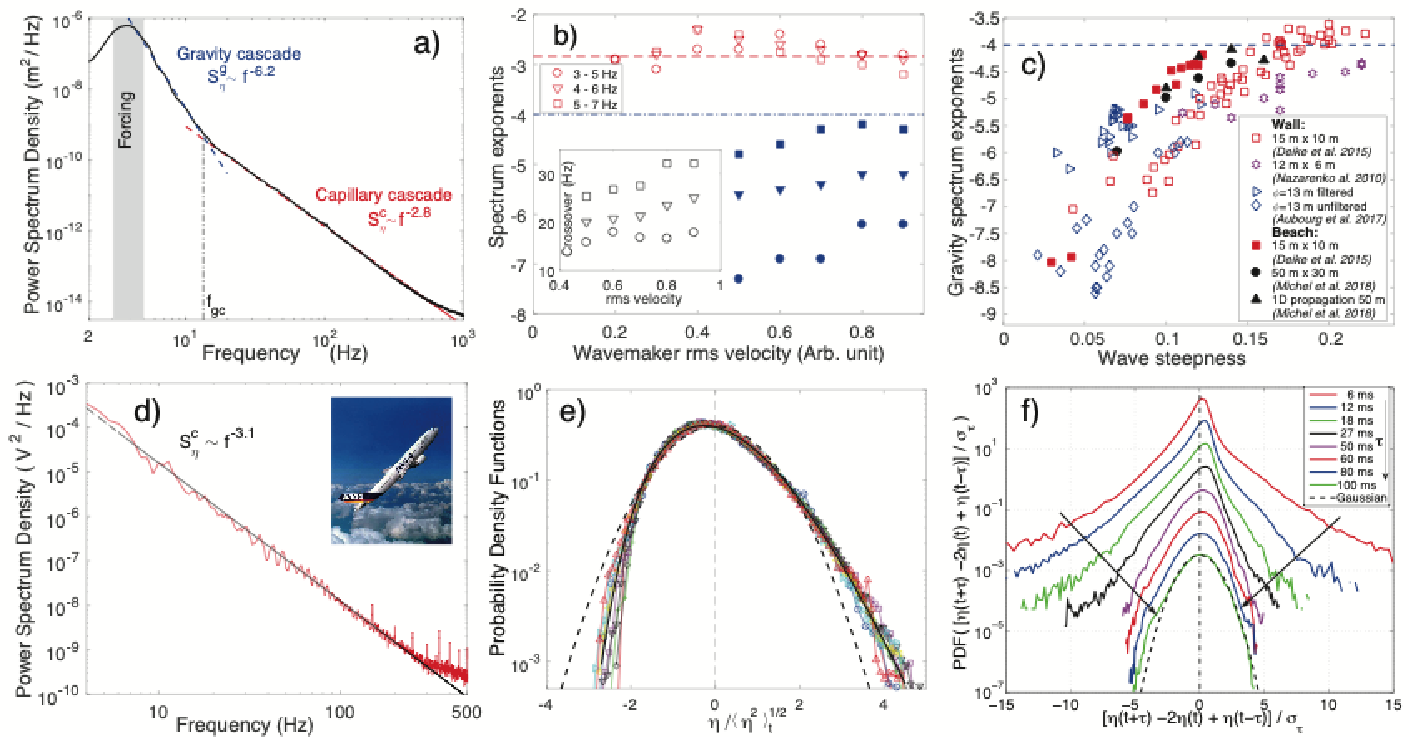}
\caption{(\textit{a}) Experimental gravity-capillary wave spectrum $S_\omega$ and best fits in the gravity and capillary regimes. $f_{gc}$ is the crossover frequency. (\textit{b}) Capillary and gravity exponents of frequency power law spectra as a function of the random forcing strength and for different forcing bandwidths. Dashed lines correspond to weak wave turbulence predictions. (\textit{c}) Wave steepness-dependent gravity exponents in different basin sizes and boundary conditions: open symbols (wall) and full symbols (beach).  (\textit{d}) Capillary-wave turbulence spectrum,  $S_\omega$, in low-gravity environment. Wave statistics: (\textit{e}) non-Gaussian distribution of dimensionless wave elevation, and (\textit{f}) distribution of the dimensionless second-order differences of wave elevation, $\eta(t+\tau)-2\eta(t)+\eta(t-\tau)$ over a time lag $\tau$, as a signature of intermittency. Panels adapted from (\textit{a}) \citet{CazaubielPRL2019} (\textit{b}) \citet{FalconPRL2007a}, (\textit{d}) \citet{FalconEPL2009}, (\textit{e}) \citet{FalconEPL11}, (\textit{f}) \citet{FalconPRL2007b} with permission of (\textit{a},\textit{b},\textit{f}) American Physical society, (\textit{d},\textit{e}) IOP Science. Panel (c) is an original creation. Inset of panel (d) courtesy of Novespace.}
\label{fig4}
\end{figure}

\section{SINGLE-POINT WAVE SPECTRUM} \label{1Dspectrum}
The wave elevation, $\eta(t)$, measured at a single point of the fluid surface, is generally found to randomly fluctuate over time. The corresponding spectrum, $S_\omega$, leads to power law scalings coexisting in the gravity and capillary regimes, for high enough nonlinearity (see \textbf{Figure \ref{fig4}a}). Each regime leads to different conclusions when compared to the predictions of Equations \ref{directg} and \ref{SpecCap}.

\subsection{Gravity Regime} \label{gravity}
In the gravity regime, the main experimental observation is that the exponent of the power law wave spectrum differs significantly from the prediction of Equation \ref{directg}, $S^g_{\omega} \sim \omega^{-4}$. The exponent was found to depend strongly on the wave steepness in experiments in laboratory basins with sizes ranging from 0.5 to 50 m \citep{FalconPRL2007a,Denissenko07,LukaschukPRL09,NazarenkoJFM2010,CobelliPRL2011,DeikeJFM2015,AubourgPRF2017} (see \textbf{Figure \ref{fig4}b,c}), as well as in field observations (see, e.g., \citet{HuangJFM1981}). The exponent depends also on the shape of the basin~\citep{IssenmannPRE13}.
The gravity spectrum was indeed found to be independent of the forcing for a cylindrical container but not for a rectangular one. 
The role of the boundary conditions (absorbing, i.e., with a beach, or reflecting, i.e., with a wall) has also been addressed \citep{DeikeJFM2015}. Although the observed stochastic wave field pattern depends strongly on these boundary conditions, their spectral properties have been found to be similar (see \textbf{Figure \ref{fig4}c}). This self-similar gravity wave spectrum (depending on the wave steepness) has been shown to be due to the presence of bound waves~\citep{MichelPRF2018,CampagnePRF2018} (see also Section~\ref{2Dspectrum}) instead of free waves. Note that for very small wave steepness ($\simeq 0.02$), the gravity spectrum has been found to be much steeper than the weak turbulence prediction, suggesting a strong impact of dissipation, although care was taken to avoid surface contamination \citep{AubourgPRF2016}. 

\begin{marginnote}[]
\entry{Free waves}{Fourier modes that follow the dispersion relation}
\entry{Bound waves}{in a stochastic wave field, bound waves are generated by nonresonant wave interactions and do not follow the linear dispersion relation} 
\end{marginnote}

In decaying wave turbulence experiments, the mean energy flux is estimated from the gravity wave energy decay rate. It is observed to be much smaller than the flux $P_b$ breaking weak turbulence theory (see Section~\ref{taunl}) \citep{BedardJETP2013,DeikeJFM2015,CazaubielPRL2019}. Nevertheless a deeper analysis of the space-time spectrum (see Section~\ref{2Dspectrum}) shows the presence of various structures associated with a finite level of nonlinearity.

The probability density functions (PDF) of wave elevation are found to be well described by the first nonlinear correction to a Gaussian distribution (i.e., a Tayfun distribution), as a confirmation of weak nonlinearity of the wave field but also of the presence of effects of a small but finite level of nonlinearity [see, e.g., \citet{FalconPRL2007a} and \textbf{Figure \ref{fig4}e}]. Moreover, intermittency in gravity-capillary wave turbulence has also been reported~\citep{FalconPRL2007b,LukaschukPRL09} (see \textbf{Figure \ref{fig4}f}). High-order differences of wave elevation need to be used when testing intermittency for signals with steep spectra as in the case of gravity-capillary waves \citep{FalconEPL10b}. This small-scale intermittency is enhanced by coherent structures (wavebreakings, capillary bursts on steep gravity waves)~\citep{FalconEPL10a} and is reduced by the wave directionality level \citep{FadaeiazarWM2018}. It also depends strongly on the forcing \citep{FalconEPL10a,DeikeJFM2015} but not on the basin boundary conditions \citep{DeikeJFM2015}.
Its origin is still an open problem and it may be related to the fractal dimension of possible singularities (e.g., peaks or wave-crest ridges) involved in the wave field \citep{Connaughton2003,NazarenkoJFM2010}. Statistics of Fourier modes (in space or in time) also reveals heavy-tail distributions attributed to the presence of large-scale coherent structures \citep{NazarenkoJFM2010}. Finally, numerical simulations of weak gravity wave turbulence have confirmed the KZ spectrum of Equation~\ref{directg} \citep{DyachenkoPRL2003,Dyachenko04}.
\begin{marginnote}
\entry{Intermittency}{when the probability density function of the wave elevation increments is Gaussian at large scales and departs strongly from being Gaussian at small scales}
\end{marginnote} 

\subsection{Capillary Regime} \label{capillary}
Capillary wave turbulence was first studied using parametric forcing and optical measurement methods \citep{Holt96,Wright1996,Wright1997,Henry2000,Lommer02,BrazhnikovEPL2002,Kolmakov2009,Snouck2009,XiaEPL2010}. This peculiar forcing led to a discrete spectrum of peaks with amplitudes decreasing as a frequency power law, since this forcing does not generate travelling waves and thus cannot be really related to kinetic wave turbulence. The use of randomly driven wave makers then led to the observation of a continuous power law wave spectrum with an exponent verifying accurately Equation~\ref{SpecCap}  ($S^c_{\omega} \sim \omega^{-17/6}$) at moderate forcing, typically $0.05 \lesssim \varepsilon \lesssim 0.15$ \citep{FalconPRL2007a,HerbertPRL2010,CobelliPRL2011,DeikePRE2012,IssenmannPRE13,DeikePRE2014}. 

For low enough viscosity, the mean energy flux scaling as $S^c_{\omega} \sim P^{1/2}$ has also been verified experimentally from the estimation of the dissipated energy spectrum \citep{DeikePRE2014}. This estimation of the mean energy flux scaling is more reliable than the one estimated from the mean injected power, which includes an unknown amount going in the bulk of the fluid \citep{FalconPRL2007a,XiaEPL2010,IssenmannPRE13}. Note also that strong temporal fluctuations of the injected power have been reported \citep{FalconPRL2008}. With the $P$- and $\omega$- scaling agreements, the KZ capillary constant can thus be estimated experimentally and is found to be one order of magnitude smaller than the theoretical one \citep{DeikePRE2014} (see Section~\ref{Solution}). This discrepancy may be ascribed to dissipation occurring at all scales of the cascade, leading to a nonconstant energy flux  \citep{DeikePRE2014}. 

The broadband dissipation is also evidenced in nonstationary wave turbulence experiments. After switching off the wave maker, the energy decay was shown to be mainly driven by the longest container eigenmodes, each Fourier mode decaying with the same damping rate \citep{DeikePRE2012,CazaubielPRL2019}. These long waves provide an energy source during the decay that sustains nonlinear interactions to keep capillary waves in a turbulent state with the expected spectrum prediction because nonlinear interactions occur faster at each scale of the cascade than dissipative processes (see e.g. \citet{CazaubielPRL2019}). 

When dissipation increases (higher viscosity fluids), the wave spectrum becomes steeper and the capillary exponent departs from its theoretical value and depends on the forcing strength, which is reminiscent of results obtained in the gravity regime \citep{DeikePRE2012}. For wave turbulence in vibrating plates, the effect of dissipation has also been unambiguously shown to steepen the spectra \citep{HumbertEPL2013,Miquel2014}. 

Pure capillary wave turbulence has been reached experimentally either in low-gravity environment [in parabolic flight experiments \citep{FalconEPL2009} or onboard the International Space Station \citep{BerhanuEPL2019}], or at the interface of two immiscible fluids of close densities, either in presence of an additional interface with air \citep{IssenmannEPL2016} or without such interface \citep{DuringPRL2009}. It leads to an excellent agreement with the $\omega^{-17/6}$ spectrum on more than two decades within the inertial range for weak enough forcing \citep{FalconEPL2009,IssenmannEPL2016} (see \textbf{Figure \ref{fig4}d}). The additional spatial symmetry in experiments of \citet{DuringPRL2009} imposes theoretically four-wave resonant interactions at the leading order, and thus a different spectrum prediction. 

Numerical simulations of isotropic weak capillary wave turbulence confirmed the KZ spectrum of Equation~\ref{SpecCap} \citep{Pushkarev1996,Pushkarev2000,PanPRL2014,DeikePRL2014,PanJFM2015}. 

Recently, quasi-1D wave capillary turbulence has been reported experimentally \citep{Ricard2021} and numerically \citep{KochurinJETP2020}. Although this geometry theoretically forbids low-order {\em resonant} interactions, a weak nonlinearity leads to the observation of unidirectional capillary-wave turbulence due to high-order resonant interactions ($N= 5$) \citep{Ricard2021}. This simple geometry should gives new perspectives in wave turbulence due to simplified calculations and measurements.

\begin{figure}[t]
\includegraphics[width=14cm]{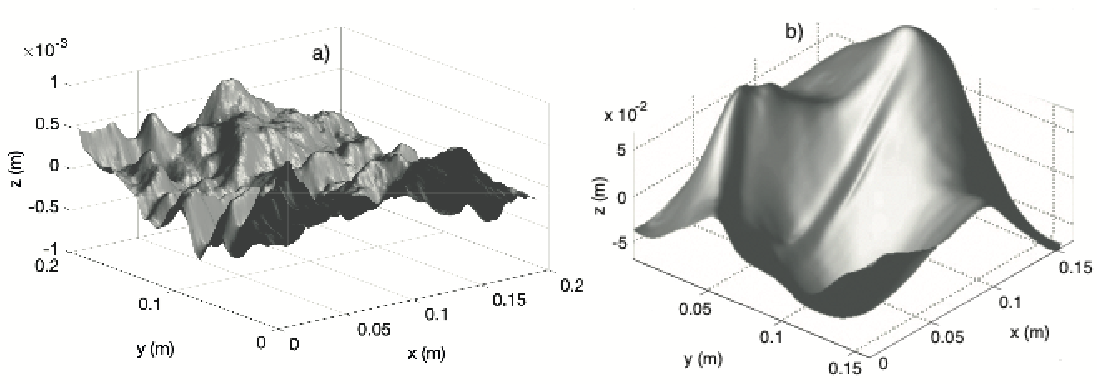} 
\caption{Wave field space-time reconstruction using the FTP method ($20\times20$ cm$^2$). (\textit{a}) Weak forcing: short capillary waves coexisting with longer gravity-capillary waves. (\textit{b}) Strong forcing: steep long waves with sharp crest ridges (coherent structures). Smaller gravity-capillary waves are not visible (the vertical scale is 50 times larger in panel \textit{b} than in panel \textit{a}). Panels adapted from (\textit{a}) \citet{AubourgPRF2017} and (\textit{b}) \citet{HerbertPRL2010} with permission of (\textit{a},\textit{b}) American Physical society.}
\label{fig5}
\end{figure}
\begin{figure}[t]
\includegraphics[width=16cm]{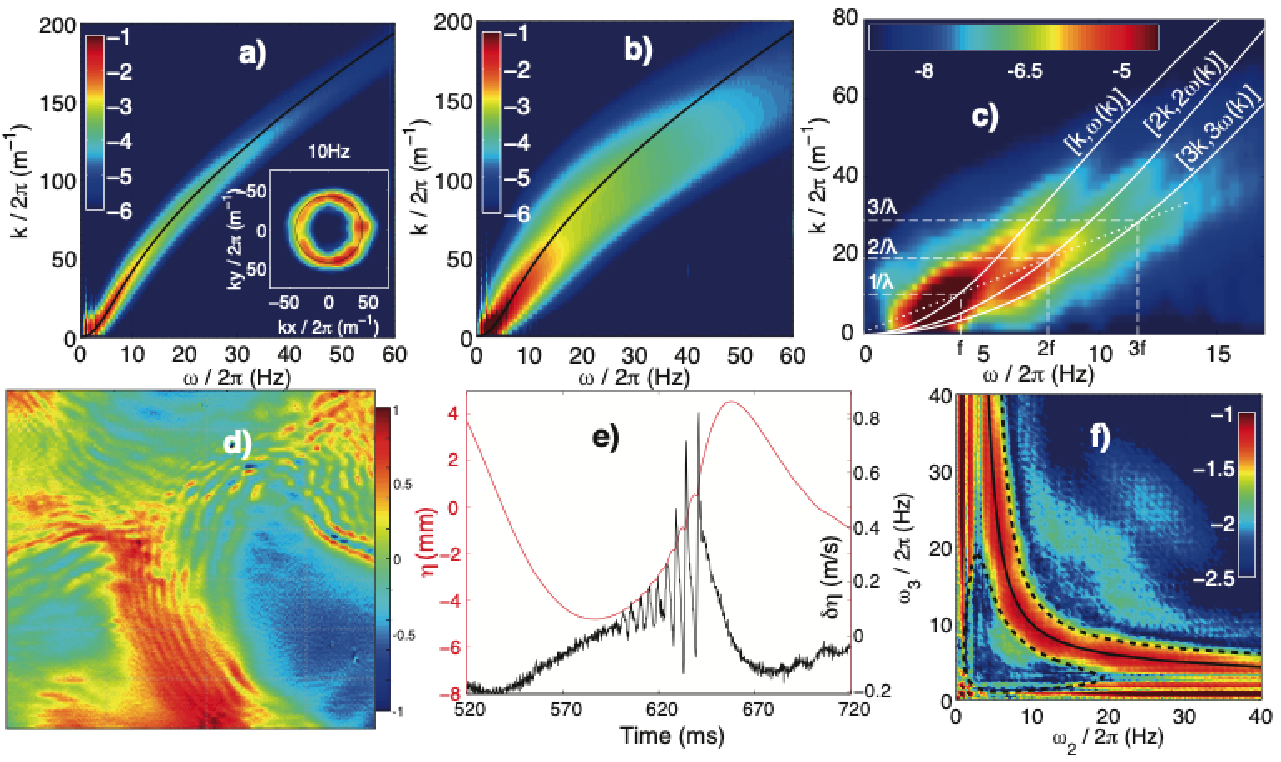} 
\caption{The experimental nonlinear dispersion relation $k/(2\pi)$ versus $\omega/(2\pi)$. (\textit{a}) Weak forcing. Inset: Wave field isotropy at fixed frequency 10 Hz. (\textit{b}) Intermediate forcing: widening of the dispersion relation. (\textit{c}) Strong forcing: an enlargement showing several branches in the dispersion relation as a signature of bound waves [$nk$,$n\omega(k)$] with $n$ an integer. (\textit{d}) The wave field $\eta(x,y)$ ($25\times25$ cm$^2$) showing gravity-capillary wave generation on steep longer waves. (\textit{e}) The same as panel (\textit{d}), recorded at a single location: wave amplitude $\eta(t)$ (black; left axis) and the corresponding wave gradient (gray; right axis). (\textit{f}) Bicoherence. The solid line indicates three-wave {\em resonant} solutions of Equation (\ref{resonance}). Dotted lines indicates boundaries of authorized three-wave {\em quasi-resonance}, i.e., solutions of approximated Equation (\ref{resonance}) affected by a nonlinear spectral spreading. Panels adapted from (\textit{a},\textit{f}) \citet{AubourgPRL2015}, (\textit{b}) \citet{AubourgPRF2016}, (\textit{c}) \citet{HerbertPRL2010}, (\textit{d}) courtesy from P. Cobelli, and (\textit{e}) \citet{FalconEPL10a} with permission of (\textit{a}-\textit{c},\textit{f}) American Physical society, (e) IOP Science.}
\label{fig6}
\end{figure}

\section{SPACE-TIME WAVE SPECTRUM} \label{2Dspectrum}
Simultaneous space-and-time-resolved measurements have been possible for a decade (see \textbf{Figure \ref{fig5}}) \citep{HerbertPRL2010} and provide a major technical improvement and a significant step forward in the understanding of wave turbulence. Spatiotemporal measurements  (see Section~\ref{spacetime}) are now able to reveal the nonlinear dispersion relation, the homogeneity of the wave field, the role of finite amplitude on the wave resonant interactions, and the role of coherent structures (bound waves, parasitic waves, sharp-crested waves, etc.) on wave turbulence. 

\textbf{Figure \ref{fig6}a-c} displays the full space-time Fourier power spectrum $S(|\mathbf{k}|,\omega)$ of the gravity-capillary wave elevation for different nonlinearity levels. At weak forcing, most of the energy injected at low $k$ is transferred to high $k$ following the linear dispersion relation (\textbf{Figure \ref{fig6}a}). Isotropy in $\mathbf{k}$-space is well verified at a given frequency in the inertial range (inset of \textbf{Figure \ref{fig6}a}), although the forcing at low frequency is often strongly anisotropic. At intermediate forcing, a nonlinear broadening of the energy distribution around the dispersion relation clearly occurs (\textbf{Figure~\ref{fig6}b}), thus authorizing numerous quasi-resonant wave interactions. The width of the nonlinear dispersion relation is a measurement of the nonlinear timescale (see Section~\ref{taunl}). At strong enough forcing, additional branches can appear as a consequence of bound waves (\textbf{Figure \ref{fig6}c})~\citep{HerbertPRL2010,MichelPRF2018,CampagnePRF2018}. The most visible ones are harmonics [$nk$,$n\omega(k)$] that propagate with the same velocity than a longer carrier wave [$k$,$\omega(k)$] as shown by the constant phase velocity in (\textbf{Figure \ref{fig6}c}).  The number $n$ of branches depends on the power injected within the waves. These coherent structures occur mainly in the gravity regime, whereas no bound waves are reported in the capillary regime. This may be due to the fact that bound waves result from nonresonant interactions at the leading order in $\varepsilon$, while weak gravity turbulence results from resonant interactions at the next order. Bound waves may contribute much less in the capillary regime since they occur at the same leading order as the numerous resonant wave interactions. In the gravity-capillary regime, capillary bursts (``parasitic waves'') are routinely observed near the crests of steep gravity-capillary waves (see \textbf{Figure \ref{fig6}e-d} and Section~\ref{taunl}). Both coherent structures lead to nonlocal energy transfer. The departure from the predictions of gravity wave turbulence, observed in field observations and in most well-controlled experiments, is thus most likely related to the spectral signature of these bound waves or other nonlinear coherent structures (see Section~\ref{gravity}).

At large levels of nonlinearity ($0.15 \lesssim \varepsilon \lesssim 0.35$), in the capillary range, the spectrum scalings in $f$ and in $k$ are surprisingly robust \citep{Berhanu2013,BerhanuJFM2018} and remain close to the KZ prediction, although the steepness is far from the weak turbulence validity range since various finite-amplitude effects are present. The space-time measurements show that wave field homogeneity and isotropy are not verified and that there is a nonlinear shift to the linear dispersion relation  \citep{Berhanu2013} that is larger than Stokes' \citep{Whitham} and Crappers' corrections \citep{Crapper}. Second, strong temporal fluctuations of the spatial spectrum have also been reported, which show stochastic bursts that transfer energy quasi-instantaneously toward small spatial scales \citep{Berhanu2013} due to the generation of parasitic capillary waves arising from a nonlocal transfer \citep{FedorovPOF1998}. To what extent this effect is related to small-scale intermittency is an open question \citep{FalconEPL10a}. 

Beyond computing the wave spectrum (second-order correlation), higher-order correlation analyses are usually performed to  quantitatively estimate the influence of three- and four-wave interactions. To verify the existence of three-wave interactions (resonant and quasi-resonant ones), researchers have computed the normalized third-order correlation $\sim |\langle \eta_{\omega_1}\eta_{\omega_2}\eta^*_{\omega_1+\omega_2}\rangle_t|$, also called bicoherence \citep{PunzmannPRL2009,AubourgPRF2016}, as shown in (\textbf{Figures~\ref{fig6}f}). The term $\eta_{\omega_1}$ is the Fourier transform of the wave elevation at $\omega_1$, $\langle \cdot \rangle_t$ denotes averaging in time, and $\ast$ is the complex conjugate. Similar correlations can be computed in wavevector space as well. It has been shown that collinear three-wave resonant interactions are dominant in the gravity-capillary regime as a result of the nonmonotonic dispersion relation \citep{AubourgPRL2015,AubourgPRF2016}. It thus becomes possible to quantify the ratio between the number of resonant waves and of quasi-resonant interactions \citep{KochurinJETP2020}. To highlight four-wave interactions, researchers have also computed the fourth-order correlation of the wave height Fourier transform $\sim |\langle\eta^*_{\omega_1}\eta^*_{\omega_2}\eta_{\omega_3}\eta_{\omega_1+\omega_2 - \omega_3}\rangle_t|$, called tricoherence \citep{AubourgPRF2017,CampagnePRF2019}. Notably, it has been shown that the four-wave quasi-resonant interactions are dominated by bound waves in the gravity regime, and can be thus filtered out to better describe their role in wave turbulence \citep{CampagnePRF2019}.

An intermediate level of nonlinearity generates coherent structures in the wave field (such as bound waves) that modify the gravity wave spectrum. At higher nonlinearity, no Phillips' spectrum has been reported experimentally so far. This is probably due to the interplay of numerous effects, including dissipation, nonlinearity and the coexistence of gravity and capillary waves. Such a strongly nonlinear regime was nevertheless observed in another system involving mechanical plates, in which a set of coherent structures coexists with weak turbulence \citep{Miquel2013}. For gravity-capillary wave turbulence, a transition near $f_{gc}$ was clearly observed on the wave spectrum separating the gravity range and the capillary range (see e.g. \textbf{Figure \ref{fig4}a}). This transition is expected from weak turbulence to change with the forcing, i.e., with $P$ (see Section~\ref{Solution}). Experimentally, the transition was found to be dependent \citep{FalconPRL2007a} (see \textbf{Figure \ref{fig4}b}) or independent \citep{IssenmannPRE13,CazaubielPRL2019} of the forcing according to whether the exponent of the gravity spectrum is dependent (as reflecting the effect of bound waves) or independent of the forcing, respectively.

The role of confinement on gravity-capillary wave turbulence has been tested by continuously decreasing a lateral dimension of a rectangular tank and keeping the other one constant \citep{HassainiPRF2018}. A discretization of the spectrum of waves propagating in this lateral direction was observed but the wave spectrum remained continuous in the unconfined direction. 
The difference between the confined and unconfined directions highlights the specificity of gravity-capillary waves in that unidirectional three-wave resonant interactions are possible (which is not the case for pure capillary waves).

The role of dispersion on gravity-capillary wave turbulence has also been investigated by decreasing the fluid depth. Finite-depth effects lead to the formation a depth-dependent hump in the capillary spectrum that could be interpreted as an analog of a bottleneck effect due to the nonlinear timescale dependence on the fluid depth \citep{FalconEPL11}. A transition has also been observed from a wave turbulence regime to a solitonic (nondispersive) regime when the forcing increases at weak depth \citep{HassainiPRF2017}. This so-called ``soliton gas'' regime is predicted by integrable turbulence \citep{Zakharov1971}, a state basically different from wave turbulence that involves coherent structures (solitons) within stochastic waves \citep{CazaubielPRF2018,RedorPRL2019,Redor2020,SuretPRL2020}. Rather than decreasing the fluid depth, the transition from a dispersive to a nondispersive regime of hydrodynamic wave turbulence can be reached using a magnetic fluid (ferrofluid) within a magnetic field \citep{BoyerPRL08,DorboloPRE11,Ricard2021b}.

\section{LARGE-SCALE WAVE TURBULENCE} \label{largescales}
The large-scale properties (i.e., larger than the forcing scale) in wave turbulence have been much less investigated experimentally, although their understanding is of primary interest (e.g., for climate modeling and long-term weather forecasting). Strong differences are expected between systems that support an inverse cascade (such as gravity waves) and systems that do not (such as pure capillary waves).

For capillary waves, since no action inverse flux is expected to large scales, their statistics are predicted to follow statistical equilibrium \citep{BalkovskyPRE1995} (see Section~\ref{zerofluxsolution}). This state was recently observed by \citet{MichelPRL2017} in capillary wave turbulence using low-viscosity fluids, whereas \citet{AbdurahimovPRE2015} found another cascade in place of the expected equilibrium state because of large-scale dissipation \citep{LvovEPL2015}.

For deep-water gravity waves, the inverse cascade spectrum of Equation~\ref{inverse} has been confirmed by direct numerical simulations \citep{AnnenkovPRL06,KorotkevitchPRL2008,KorotkevitchMCS2008}. Laboratory observations of the inverse cascade in gravity wave turbulence were limited to a narrow inertial range due to the small size of the container used~\citep{DeikeEPL2011}, while recent attempts have been inconclusive using a larger cylindrical basin \citep{CampagneProc2019} or another type of forcing \citep{NazarenkoRev2016}. \citet{MichelPRF2019} showed that in a cylindrical container, gravity waves may sustain three-wave resonant interactions due to confinement as a consequence of a new conserved quantity (angular pseudo-momentum). These interactions have been evidenced in a gravity wave turbulence experiment in a high-gravity environment \citep{CazaubielPRL2019}. Thus, the absence of an inverse cascade in \citet{CampagneProc2019} study may be due to finite-size effects due to the specific shape used. By replacing the usual absorbing beach of a large-scale rectangular wave basin by a reflective wall and by forcing multidirectional random waves to foster wave interactions and homogeneity, \citet{FalconPRL2020} observed an inverse cascade of gravity wave turbulence, compatible with the prediction of weak turbulence theory. However, the inverse cascade was found to stop well before a condensate state is reached, i.e., before wave action piles up at the largest scale due to basin finite size effects. The limitation of the inverse cascade results from the emergence of dissipative coherent structures (sharp-crested waves) that are necessary in order to balance fluxes and allow the system to reach a statistically stationary state \citep{FalconPRL2020}.

\begin{summary}[SUMMARY POINTS]
\begin{enumerate}
\item The basic mechanism of energy transfer in weak turbulence theory (wave resonant interactions) is validated experimentally in the gravity (four-wave interactions) and capillary (three-wave interactions) regimes. However, dissipation and nonlinearity broaden the dispersion relation and thus authorize more quasi-resonant interactions than just resonant ones. Among them, unidirectional resonant interactions occur near the crossover, as the gravity-capillary dispersion relation is not a pure power law.
\item Experimental capillary wave turbulence is well described by weak turbulence theory for weak enough nonlinearity, although broad-scale dissipation must be taken into account. On the contrary, the gravity wave spectrum is in strong disagreement because of the presence of bound waves.
\item Boundary conditions and finite-size effects play a significant role in gravity-capillary wave turbulence, and their effect is beginning to be considered.
\item Finite-amplitude effects have a crucial role in experiments in that they regularize fluxes by means of coherent structures (e.g., sharp crested waves, parasitic waves, bound waves, wave breakings) involving nonlocal interactions, sinks of dissipation, and energy flux fluctuations that are not taken into account theoretically.
\item Thanks to advances in experiments, it is now possible to achieve full spatiotemporal reconstruction of the wave field in a weakly or strong nonlinear regime in order to infer wave statistics, waves' nonlinear dispersion relationship, and the wave-field homogeneity level so that these can be accurately compared with theoretical predictions.
\end{enumerate}
\end{summary}

\begin{issues}[FUTURE ISSUES]
\begin{enumerate}
\item The role of finite-size effects on wave resonant interactions and on wave turbulence merits further studies with accurate space-time measurements.
\item Investigating the role of strongly nonlinear waves (coherent structures, bound waves, sharp-crested crests, parasitic waves, etc.) and the corresponding nonlocal interactions in strong wave turbulence would facilitate better modeling of field observations.
\item Performing direct numerical simulations of gravity-capillary wave turbulence including dissipation is necessary for comparison with experiments and theories.
\item Extending the theory to more realistic conditions (e.g., broadband dissipation, finite nonlinearity, finite size) would help researchers to understand observations and to develop numerical codes for sea state forecasting. 
\item Better identifying the mechanisms (such as inverse cascade, statistical equilibrium, condensate) governing large-scale properties of gravity-capillary wave turbulence is of paramount interest, notably for climat modeling.
\end{enumerate}
\end{issues}

\section*{DISCLOSURE STATEMENT}
The authors are not aware of any affiliations, memberships, funding, or financial holdings that might be perceived as affecting the objectivity of this review. 

\section*{ACKNOWLEDGMENTS}
This work was supported by the Simons Foundation (MPS grant 651463 Wave Turbulence). E.F. was also supported by the French National Research Agency (ANR) (grants DYSTURB ANR-17-CE30-0004, TURBULON ANR-12-BS04-0005, and TURBONDE ANR-07-BLAN-0246), and N.M. by the European Research Council (grant 647018-WATU). We acknowledge all our co-authors listed in the references below. We thank P. Cobelli for Figures \ref{fig3}b and \ref{fig6}d, J.-B. Gorce for Figure \ref{fig3}d, Novespace for inset of Figure \ref{fig4}d, and G. Michel and G. Ricard for proofreading.

%
\bibliography{biblioARFMviaPapers3} 

\begin{thebibliography}{}
\expandafter\ifx\csname natexlab\endcsname\relax\def\natexlab#1{#1}\fi

\bibitem[Abdurakhimov et~al.(2015)Abdurakhimov, Arefin, Kolmakov, Levchenko,
  L'Vov \& Remizov]{AbdurahimovPRE2015}
Abdurakhimov LV, Arefin M, Kolmakov GV, Levchenko AA, L'Vov YV, Remizov IA.
  2015.
{Bidirectional energy cascade in surface capillary waves}.
\textit{Phys. Rev. E} 91:023021

\bibitem[Abella \& Soriano(2019)]{AbellaPS2019}
Abella AP, Soriano MN. 2019.
{Detection and visualization of water surface three-wave resonance via
  synthetic Schlieren method}.
\textit{Phys. Scr.} 94:034006

\bibitem[Annenkov \& Shrira(2006)]{AnnenkovPRL06}
Annenkov SY, Shrira VI. 2006.
{Direct numerical simulation of downshift and inverse cascade for water wave
  turbulence}.
\textit{Phys. Rev. Lett.} 96:204501

\bibitem[Aubourg et~al.(2017)Aubourg, Campagne, Peureux, Ardhuin, Sommeria
  et~al.]{AubourgPRF2017}
Aubourg Q, Campagne A, Peureux C, Ardhuin F, Sommeria J, et~al. 2017.
{Three-wave and four-wave interactions in gravity wave turbulence}.
\textit{Phys. Rev. Fluids.} 2:114802

\bibitem[Aubourg \& Mordant(2015)]{AubourgPRL2015}
Aubourg Q, Mordant N. 2015.
{Nonlocal resonances in weak turbulence of gravity-capillary waves}.
\textit{Phys. Rev. Lett.} 114:144501

\bibitem[Aubourg \& Mordant(2016)]{AubourgPRF2016}
Aubourg Q, Mordant N. 2016.
{Investigation of resonances in gravity-capillary wave turbulence}.
\textit{Phys. Rev. Fluids.} 1:023701

\bibitem[Balkovsky et~al.(1995)Balkovsky, Falkovich, Lebedev \&
  Shapiro]{BalkovskyPRE1995}
Balkovsky E, Falkovich G, Lebedev V, Shapiro IY. 1995.
{Large-scale properties of wave turbulence}.
\textit{Phys. Rev. E} 52:4537

\bibitem[Bedard et~al.(2013)Bedard, Lukaschuk \& Nazarenko]{BedardJETP2013}
Bedard R, Lukaschuk S, Nazarenko S. 2013.
{Non-stationary regimes of surface gravity wave turbulence}.
\textit{JETP Lett.} 97:459--465

\bibitem[Benetazzo(2006)]{Benetazzo2006}
Benetazzo A. 2006.
Measurements of short water waves using stereo matched image sequences.
\textit{Coast. Engin.} 53:1013 -- 1032

\bibitem[Berhanu \& Falcon(2013)]{Berhanu2013}
Berhanu M, Falcon E. 2013.
{Space-time resolved capillary wave turbulence}.
\textit{Phys. Rev. E} 87:033003

\bibitem[Berhanu et~al.(2018)Berhanu, Falcon \& Deike]{BerhanuJFM2018}
Berhanu M, Falcon E, Deike L. 2018.
Turbulence of capillary waves forced by steep gravity waves.
\textit{J. Fluid Mech.} 850:803--843

\bibitem[Berhanu et~al.(2019)Berhanu, Falcon, Michel, Gissinger \&
  Fauve]{BerhanuEPL2019}
Berhanu M, Falcon E, Michel G, Gissinger C, Fauve S. 2019.
{Capillary wave turbulence experiments in microgravity}.
\textit{Europhys. Lett.} 128:34001

\bibitem[Bonnefoy et~al.(2016)Bonnefoy, Haudin, Michel, Semin, Humbert
  et~al.]{BonnefoyJFM2016}
Bonnefoy F, Haudin F, Michel G, Semin B, Humbert T, et~al. 2016.
{Observation of resonant interactions among surface gravity waves}.
\textit{J. Fluid. Mech.} 805:R3

\bibitem[Bonnefoy et~al.(2017)Bonnefoy, Haudin, Michel, Semin, Humbert
  et~al.]{BonnefoyHouille17}
Bonnefoy F, Haudin F, Michel G, Semin B, Humbert T, et~al. 2017.
{Experimental observation of four-wave resonant interactions in a wave basin}.
\textit{La Houille Blanche} 5:56.
(in french)

\bibitem[Boyer \& Falcon(2008)]{BoyerPRL08}
Boyer F, Falcon E. 2008.
{Wave turbulence on the surface of a ferrofluid in a magnetic field}.
\textit{Phys. Rev. Lett.} 101:244502

\bibitem[Brazhnikov et~al.(2002)Brazhnikov, Kolmakov, Levchenko \&
  Mezhov-Deglin]{BrazhnikovEPL2002}
Brazhnikov MY, Kolmakov GV, Levchenko AA, Mezhov-Deglin LP. 2002.
{Observation of capillary turbulence on the water surface in a wide range of
  frequencies}.
\textit{Europhys. Lett.} 58:510

\bibitem[Cabrera \& Cobelli(2021)]{Cabrera2021}
Cabrera F, Cobelli PJ. 2021.
Design, construction and validation of an instrumented particle for the
  lagrangian characterization of flows: application to gravity wave turbulence

\bibitem[Campagne et~al.(2019{\natexlab{a}})Campagne, Hassaini, Redor, Sommeria
  \& Mordant]{CampagneProc2019}
Campagne A, Hassaini R, Redor I, Sommeria J, Mordant N. 2019{\natexlab{a}}.
The energy cascade of surface wave turbulence: Toward identifying the active
  wave coupling, In \textit{Turbulent Cascades II}, eds. M~Gorokhovski,
  FS~Godeferd,\ pp.  239--246, Cham: Springer International Publishing

\bibitem[Campagne et~al.(2018)Campagne, Hassaini, Redor, Sommeria, Valran
  et~al.]{CampagnePRF2018}
Campagne A, Hassaini R, Redor I, Sommeria J, Valran T, et~al. 2018.
{Impact of dissipation on the energy spectrum of experimental turbulence of
  gravity surface waves}.
\textit{Phys. Rev. Fluids} 3:044801

\bibitem[Campagne et~al.(2019{\natexlab{b}})Campagne, Hassaini, Redor, Valran,
  Viboud et~al.]{CampagnePRF2019}
Campagne A, Hassaini R, Redor I, Valran T, Viboud S, et~al. 2019{\natexlab{b}}.
{Identifying four-wave-resonant interactions in a surface gravity wave
  turbulence experiment}.
\textit{Phys. Rev. Fluids} 7:074801

\bibitem[Cazaubiel et~al.(2019{\natexlab{a}})Cazaubiel, Haudin, Falcon \&
  Berhanu]{CazaubielPRF2019}
Cazaubiel A, Haudin F, Falcon E, Berhanu M. 2019{\natexlab{a}}.
{Forced three-wave interactions of gravity-capillary surface waves}.
\textit{Phys. Rev. Fluids} 4:074803

\bibitem[Cazaubiel et~al.(2019{\natexlab{b}})Cazaubiel, Mawet, Darras, Grojean,
  van Loon et~al.]{CazaubielPRL2019}
Cazaubiel A, Mawet S, Darras A, Grojean G, van Loon JJWA, et~al.
  2019{\natexlab{b}}.
{Wave turbulence on the surface of a fluid in a high-gravity environment}.
\textit{Phys. Rev. Lett.} 123:244501

\bibitem[Cazaubiel et~al.(2018)Cazaubiel, Michel, Lepot, Semin, Auma\^{\i}tre
  et~al.]{CazaubielPRF2018}
Cazaubiel A, Michel G, Lepot S, Semin B, Auma\^{\i}tre S, et~al. 2018.
Coexistence of solitons and extreme events in deep water surface waves.
\textit{Phys. Rev. Fluids} 3:114802

\bibitem[Cobelli et~al.(2009{\natexlab{a}})Cobelli, Maurel, Pagneux \&
  Petitjeans]{CobelliEF2009}
Cobelli P, Maurel A, Pagneux V, Petitjeans P. 2009{\natexlab{a}}.
{Global measurement of water waves by Fourier transform profilometry,}.
\textit{Exp Fluids} 46:1037

\bibitem[Cobelli et~al.(2009{\natexlab{b}})Cobelli, Petitjeans, Maurel, Pagneux
  \& Mordant]{CobelliPRL2009}
Cobelli P, Petitjeans P, Maurel A, Pagneux V, Mordant N. 2009{\natexlab{b}}.
Space-time resolved wave turbulence in a vibrating plate.
\textit{Phys. Rev. Lett.} 103:204301

\bibitem[Cobelli et~al.(2011)Cobelli, Przadka, Petitjeans, Lagubeau, Pagneux \&
  Maurel]{CobelliPRL2011}
Cobelli P, Przadka A, Petitjeans P, Lagubeau G, Pagneux V, Maurel A. 2011.
{Different regimes for water wave turbulence}.
\textit{Phys. Rev. Lett.} 107:214503

\bibitem[Connaughton et~al.(2003)Connaughton, Nazarenko \&
  Newell]{Connaughton2003}
Connaughton C, Nazarenko S, Newell AC. 2003.
{Dimensional analysis and weak turbulence}.
\textit{Physica D} 184:86--97

\bibitem[Crapper(1957)]{Crapper}
Crapper GD. 1957.
{An exact solution for progressive capillary waves of arbitrary amplitude}.
\textit{J. Fluids Mech.} 2:532

\bibitem[Dalziel et~al.(2000)Dalziel, Hughes \& Sutherland]{Dalziel2000}
Dalziel SB, Hughes GO, Sutherland BR. 2000.
{Whole-field density measurements by ``synthetic schlieren''}.
\textit{Exp. Fluids} 28:322--335

\bibitem[Davis et~al.(2020)Davis, Jamin, Deleuze, Joubaud \&
  Dauxois]{DavisPRL2020}
Davis G, Jamin T, Deleuze J, Joubaud S, Dauxois T. 2020.
{Succession of resonances to achieve internal wave turbulence}.
\textit{Phys. Rev. Lett.} 124:204502

\bibitem[Deike et~al.(2013)Deike, Bacri \& Falcon]{DeikeJFM2013}
Deike L, Bacri JC, Falcon E. 2013.
{Nonlinear waves on the surface of a fluid covered by an elastic sheet}.
\textit{J. Fluid. Mech.} 733:394--413

\bibitem[Deike et~al.(2012)Deike, Berhanu \& Falcon]{DeikePRE2012}
Deike L, Berhanu M, Falcon E. 2012.
{Decay of capillary wave turbulence}.
\textit{Phys. Rev. E} 85:066311

\bibitem[Deike et~al.(2014{\natexlab{a}})Deike, Berhanu \&
  Falcon]{DeikePRE2014}
Deike L, Berhanu M, Falcon E. 2014{\natexlab{a}}.
{Energy flux measurement from the dissipated energy in capillary wave
  turbulence}.
\textit{Phys. Rev. E} 89:023003

\bibitem[Deike et~al.(2014{\natexlab{b}})Deike, Fuster, Berhanu \&
  Falcon]{DeikePRL2014}
Deike L, Fuster D, Berhanu M, Falcon E. 2014{\natexlab{b}}.
{Direct Numerical Simulations of Capillary Wave Turbulence}.
\textit{Phys. Rev. Lett.} 112:234501

\bibitem[Deike et~al.(2011)Deike, Laroche \& Falcon]{DeikeEPL2011}
Deike L, Laroche C, Falcon E. 2011.
{Experimental study of the inverse cascade in gravity wave turbulence}.
\textit{Europhys. Lett.} 96:34004

\bibitem[Deike et~al.(2015)Deike, Miquel, Guti{\'e}rrez, Jamin, Semin
  et~al.]{DeikeJFM2015}
Deike L, Miquel B, Guti{\'e}rrez P, Jamin T, Semin B, et~al. 2015.
{Role of the basin boundary conditions in gravity wave turbulence}.
\textit{J. Fluid. Mech.} 781:196--225

\bibitem[Del~Grosso et~al.(2019)Del~Grosso, Cappelletti, Sujovolsky, Mininni \&
  Cobelli]{DelGrossoPRF2019}
Del~Grosso NF, Cappelletti LM, Sujovolsky NE, Mininni PD, Cobelli PJ. 2019.
Statistics of single and multiple floaters in experiments of surface wave
  turbulence.
\textit{Phys. Rev. Fluids} 4:074805

\bibitem[Denissenko et~al.(2007)Denissenko, Lukaschuk \&
  Nazarenko]{Denissenko07}
Denissenko P, Lukaschuk S, Nazarenko S. 2007.
{Gravity wave turbulence in a laboratory flume}.
\textit{Phys. Rev. Lett.} 99:014501

\bibitem[Dorbolo \& Falcon(2011)]{DorboloPRE11}
Dorbolo S, Falcon E. 2011.
{Wave turbulence on the surface of a ferrofluid in a horizontal magnetic
  field}.
\textit{Phys. Rev. E} 83:046303

\bibitem[D{\"u}ring \& Falc{\'o}n(2009)]{DuringPRL2009}
D{\"u}ring G, Falc{\'o}n C. 2009.
{Symmetry induced four-wave capillary wave turbulence}.
\textit{Phys. Rev. Lett.} 103:174503

\bibitem[Dyachenko et~al.(2004)Dyachenko, Korotkevich \& Zakharov]{Dyachenko04}
Dyachenko AI, Korotkevich A, Zakharov VE. 2004.
{Weak turbulent Kolmogorov spectrum for surface gravity waves}.
\textit{Phys. Rev. Lett.} 92:134501

\bibitem[Dyachenko et~al.(2003)Dyachenko, Korotkevich \&
  Zakharov]{DyachenkoPRL2003}
Dyachenko AI, Korotkevich AO, Zakharov VE. 2003.
{Weak turbulence of gravity waves}.
\textit{JETP Lett.} 77:546

\bibitem[Fadaeiazar et~al.(2018)Fadaeiazar, Alberello, Onorato, Leontini,
  Frascoli et~al.]{FadaeiazarWM2018}
Fadaeiazar E, Alberello A, Onorato M, Leontini J, Frascoli F, et~al. 2018.
Wave turbulence and intermittency in directional wave fields.
\textit{Wave Mot.} 83:94 -- 101

\bibitem[Falc{\'o}n et~al.(2009)Falc{\'o}n, Falcon, Bortolozzo \&
  Fauve]{FalconEPL2009}
Falc{\'o}n C, Falcon E, Bortolozzo U, Fauve S. 2009.
{Capillary wave turbulence on a spherical fluid surface in low gravity}.
\textit{Europhys. Lett.} 86:14002

\bibitem[Falcon(2010)]{FalconDCDSB2010}
Falcon E. 2010.
{Laboratory experiments on wave turbulence}.
\textit{Discrete Contin. Dyn. Syst. B} 13:819

\bibitem[Falcon(2019)]{Hawai}
Falcon E. 2019.
{Wave turbulence: a set of stochastic nonlinear waves in interaction}. In
  \textit{Proceedings of the 5th International Conference on Applications in
  Nonlinear Dynamics}, eds. V~In, P~Longhini, A~Palacios. New York: Springer,
  259--266, Chap. 25

\bibitem[Falcon et~al.(2008)Falcon, Aumaître, Falc{\'o}n, Laroche \&
  Fauve]{FalconPRL2008}
Falcon E, Aumaître S, Falc{\'o}n C, Laroche C, Fauve S. 2008.
{Fluctuations of energy flux in wave turbulence}.
\textit{Phys. Rev. Lett.} 100:064503

\bibitem[Falcon et~al.(2007{\natexlab{a}})Falcon, Fauve \&
  Laroche]{FalconPRL2007b}
Falcon E, Fauve S, Laroche C. 2007{\natexlab{a}}.
{Observation of intermittency in wave turbulence}.
\textit{Phys. Rev. Lett.} 98:154501

\bibitem[Falcon \& Laroche(2011)]{FalconEPL11}
Falcon E, Laroche C. 2011.
{Observation of depth-induced properties in wave turbulence on the surface of a
  fluid}.
\textit{Europhys. Lett.} 94:34003

\bibitem[Falcon et~al.(2003)Falcon, Laroche \& Fauve]{FalconPRL2003}
Falcon E, Laroche C, Fauve S. 2003.
{Observation of Sommerfeld precursors on a fluid surface}.
\textit{Phys. Rev. Lett.} 91:064502

\bibitem[Falcon et~al.(2007{\natexlab{b}})Falcon, Laroche \&
  Fauve]{FalconPRL2007a}
Falcon E, Laroche C, Fauve S. 2007{\natexlab{b}}.
Observation of gravity-capillary wave turbulence.
\textit{Phys. Rev. Lett.} 98:094503

\bibitem[Falcon et~al.(2020)Falcon, Michel, Prabhudesai, Cazaubiel, Berhanu
  et~al.]{FalconPRL2020}
Falcon E, Michel G, Prabhudesai G, Cazaubiel A, Berhanu M, et~al. 2020.
Saturation of the inverse cascade in surface gravity-wave turbulence.
\textit{Phys. Rev. Lett.} 125:134501

\bibitem[Falcon et~al.(2010{\natexlab{a}})Falcon, Roux \& Audit]{FalconEPL10b}
Falcon E, Roux SG, Audit B. 2010{\natexlab{a}}.
{Revealing intermittency in experimental data with steep power spectra}.
\textit{Europhys. Lett.} 90:50007

\bibitem[Falcon et~al.(2010{\natexlab{b}})Falcon, Roux \&
  Laroche]{FalconEPL10a}
Falcon E, Roux SG, Laroche C. 2010{\natexlab{b}}.
{On the origin of intermittency in wave turbulence}.
\textit{Europhys. Lett.} 90:34005

\bibitem[Fedorov et~al.(1998)Fedorov, Melville \& Rozenberg]{FedorovPOF1998}
Fedorov AV, Melville WK, Rozenberg A. 1998.
{An experimental and numerical study of parasitic capillary waves}.
\textit{Phys. Fluids} 10:1315

\bibitem[Galtier(2020)]{GaltierAFD2020}
Galtier S. 2020.
Wave turbulence: the case of capillary waves.
\textit{Geophys. Astrophys. Fluid Dyn.} :1--24

\bibitem[Hassaini \& Mordant(2017)]{HassainiPRF2017}
Hassaini R, Mordant N. 2017.
{Transition from weak wave turbulence to soliton gas}.
\textit{Phys. Rev. Fluids} 2:094803

\bibitem[Hassaini \& Mordant(2018)]{HassainiPRF2018}
Hassaini R, Mordant N. 2018.
{Confinement effects on gravity-capillary wave turbulence}.
\textit{Phys. Rev. Fluids} 3:094805

\bibitem[Hasselmann(1962)]{Hasselmann62}
Hasselmann K. 1962.
{On the non-linear energy transfer in a gravity-wave spectrum. I. General
  theory}.
\textit{J. Fluid. Mech.} 12:481--500

\bibitem[Haudin et~al.(2016)Haudin, Cazaubiel, Deike, Jamin, Falcon \&
  Berhanu]{HaudinPRE2016}
Haudin F, Cazaubiel A, Deike L, Jamin T, Falcon E, Berhanu M. 2016.
{Experimental study of three-wave interactions among capillary-gravity surface
  waves}.
\textit{Phys. Rev. E} 93:043110

\bibitem[Henderson \& Hammack(1987)]{HendersonJFM1987}
Henderson DM, Hammack JL. 1987.
{Experiments on ripple instabilities, Part 1. Resonant Triads}.
\textit{J. Fluid. Mech} 184:15--41

\bibitem[Henderson \& Miles(1990)]{HendersonJFM1990}
Henderson DM, Miles JW. 1990.
Single-mode faraday waves in small cylinders.
\textit{J. Fluid Mech.} 213:95--109

\bibitem[Henry et~al.(2000)Henry, Alstrom \& Levinsen]{Henry2000}
Henry E, Alstrom P, Levinsen MT. 2000.
{Prevalence of weak turbulence in strongly driven surface ripples}.
\textit{Europhys. Lett.} 52:27

\bibitem[Herbert et~al.(2010)Herbert, Mordant \& Falcon]{HerbertPRL2010}
Herbert E, Mordant N, Falcon E. 2010.
Observation of the nonlinear dispersion relation and spatial statistics of wave
  turbulence on the surface of a fluid.
\textit{Phys. Rev. Lett.} 105:144502

\bibitem[Holt \& Trinh(1996)]{Holt96}
Holt RG, Trinh EH. 1996.
{Faraday wave turbulence on a spherical liquid shell}.
\textit{Phys. Rev. Lett.} 77:1274--1277

\bibitem[Hrabski \& Pan(2020)]{HrabskiPRE2020}
Hrabski A, Pan Y. 2020.
Effect of discrete resonant manifold structure on discrete wave turbulence.
\textit{Phys. Rev. E} 102:041101

\bibitem[Huang et~al.(1981)Huang, Long, Tung, Yuen \& Bliven]{HuangJFM1981}
Huang NE, Long SR, Tung CC, Yuen Y, Bliven LF. 1981.
A unified two-parameter wave spectral model for a general sea state.
\textit{J. Fluid Mech.} 112:203 -- 224

\bibitem[Humbert et~al.(2013)Humbert, Cadot, D\"uring, Josserand, Rica \&
  Touz\'e]{HumbertEPL2013}
Humbert T, Cadot O, D\"uring G, Josserand C, Rica S, Touz\'e C. 2013.
Wave turbulence in vibrating plates : the effect of damping.
\textit{Europhys. Lett. (EPL)} 102:30002

\bibitem[Issenmann \& Falcon(2013)]{IssenmannPRE13}
Issenmann B, Falcon E. 2013.
{Gravity wave turbulence revealed by horizontal vibrations of the container}.
\textit{Phys. Rev. E} 87:011001(R)

\bibitem[Issenmann et~al.(2016)Issenmann, Laroche \& Falcon]{IssenmannEPL2016}
Issenmann B, Laroche C, Falcon E. 2016.
{Wave turbulence in a two-layer fluid: coupling between free surface and
  interface waves}.
\textit{Europhys. Lett. (EPL)} 116:64005

\bibitem[Kartashova et~al.(2008)Kartashova, Nazarenko \&
  Rudenko]{KartashovaPRE08}
Kartashova E, Nazarenko S, Rudenko O. 2008.
{Resonant interactions of nonlinear water waves in a finite basin}.
\textit{Phys. Rev. E} 78:016304

\bibitem[Kityk et~al.(2004)Kityk, Knorr, Müller \& Wagner]{Kityk2004}
Kityk AV, Knorr K, Müller HW, Wagner C. 2004.
Spatio-temporal fourier analysis of faraday surface wave patterns on a
  two-liquid interface.
\textit{Europhys. Lett.} 65:857--863

\bibitem[Kochurin et~al.(2020)Kochurin, Ricard, Zubarev \&
  Falcon]{KochurinJETP2020}
Kochurin E, Ricard G, Zubarev N, Falcon E. 2020.
Numerical simulation of collinear capillary-wave turbulence.
\textit{JETP Lett.} 112:757--763

\bibitem[Kolmakov et~al.(2009)Kolmakov, Brazhnikov, Levchenko, Abdurakhimov,
  McClintock \& Mezhov-Deglin]{Kolmakov2009}
Kolmakov GV, Brazhnikov MY, Levchenko AA, Abdurakhimov LV, McClintock PVE,
  Mezhov-Deglin LP. 2009.
{Capillary Turbulence on the Surfaces of Quantum Fluids}. In \textit{Progress
  inLow Temperature Physics: Quantum Turbulence}, eds. M~Tsubota, WP~Halperin.
  Elsevier,  305--349

\bibitem[Korotkevich(2012)]{KorotkevitchMCS2008}
Korotkevich A. 2012.
{Influence of the condensate and inverse cascade on the direct cascade in wave
  turbulence}.
\textit{Math. Comput. Simul.} 82:1228

\bibitem[Korotkevitch(2008)]{KorotkevitchPRL2008}
Korotkevitch AO. 2008.
Simultaneous numerical simulation of direct and inverse cascades in wave
  turbulence.
\textit{Phys. Rev. Lett.} 101:074501

\bibitem[Kraichnan(1965)]{KraichnanPOF1965}
Kraichnan R. 1965.
{Inertial range spectrum of hydro-magnetic turbulence}.
\textit{Phys. Fluids} 8:1385–1387

\bibitem[Kurata et~al.(1990)Kurata, Grattan, Uchiyama \& Tanaka]{Kurata1990}
Kurata J, Grattan KTV, Uchiyama H, Tanaka T. 1990.
Water surface measurement in a shallow channel using the transmitted image of a
  grating.
\textit{Rev. Sci. Instr.} 61:736--739

\bibitem[Kuznetsov(2004)]{Kuznetsov04}
Kuznetsov EA. 2004.
{Turbulence spectra generated by singularities}.
\textit{JETP Lett.} 80:83--89

\bibitem[Lamb(1932)]{Lamb1932}
Lamb H. 1932.
\em{Hydrodynamics}.
Springer-Verlag, Berlin

\bibitem[Leckler et~al.(2015)Leckler, Ardhuin, Peureux, Benetazzo, Bergamasco
  \& Dulov]{Leckler15}
Leckler F, Ardhuin F, Peureux C, Benetazzo A, Bergamasco F, Dulov V. 2015.
{Analysis and interpretation of frequency-wavenumber spectra of young wind
  waves}.
\textit{J. Phys. Ocean.} 45:2484

\bibitem[Lommer \& Levinsen(2002)]{Lommer02}
Lommer M, Levinsen MT. 2002.
{Using laser-induced fluorescence in the study of surface wave turbulence}.
\textit{J. Fluoresc.} 12:45--50

\bibitem[Longuet-Higgins(1962)]{Longuet-Higgins1962}
Longuet-Higgins MS. 1962.
{Resonant interactions between two trains of gravity waves}.
\textit{J. Fluid. Mech.} 12:321--332

\bibitem[Longuet-Higgins(1963)]{LonguetHiggins1963}
Longuet-Higgins MS. 1963.
The generation of capillary waves by steep gravity waves.
\textit{J. Fluid. Mech.} 16:138--159

\bibitem[Longuet-Higgins \& Smith(1966)]{Longuet-Higgins1966}
Longuet-Higgins MS, Smith ND. 1966.
{An experiment on third-order resonant wave interactions.}
\textit{J. Fluid. Mech.} 25:417--435

\bibitem[Lukaschuk et~al.(2009)Lukaschuk, Nazarenko, McLelland \&
  Denissenko]{LukaschukPRL09}
Lukaschuk S, Nazarenko S, McLelland S, Denissenko P. 2009.
Gravity wave turbulence in wave tanks: Space and time statistics.
\textit{Phys. Rev. Lett.} 103:044501

\bibitem[L'vov \& Nazarenko(2010)]{LvovPRE10}
L'vov VS, Nazarenko S. 2010.
Discrete and mesoscopic regimes of finite-size wave turbulence.
\textit{Phys. Rev. E} 82:056322

\bibitem[Lvov et~al.(2015)Lvov, He \& Kolmakov]{LvovEPL2015}
Lvov YV, He A, Kolmakov GV. 2015.
Formation of the bi-directional energy cascade in low-frequency damped
  wave-turbulent systems.
\textit{Europhys. Lett.} 112:24004

\bibitem[McGoldrick(1965)]{McGoldrick1965}
McGoldrick LF. 1965.
{Resonant interactions among capillary-gravity waves}.
\textit{J. Fluid. Mech.} 21:305

\bibitem[McGoldrick(1970)]{McGoldrick1970}
McGoldrick LF. 1970.
{An experiment on second-order capillary gravity resonant wave interactions}.
\textit{J. Fluid Mech.} 40:251

\bibitem[McGoldrick et~al.(1966)McGoldrick, Phillips, Huang \&
  Hodgson]{McGoldrick1966}
McGoldrick LF, Phillips O, Huang NE, Hodgson TH. 1966.
{Measurements of third-order resonant wave interactions}.
\textit{J. Fluid Mech.} 25:437--456

\bibitem[Michel(2019)]{MichelPRF2019}
Michel G. 2019.
{Three-wave interactions among surface gravity waves in a cylindrical
  container}.
\textit{Phys. Rev. Fluids} 4:012801(R)

\bibitem[Michel et~al.(2017)Michel, P{\'e}tr{\'e}lis \& Fauve]{MichelPRL2017}
Michel G, P{\'e}tr{\'e}lis F, Fauve S. 2017.
Observation of thermal equilibrium in capillary wave turbulence.
\textit{Phys. Rev. Lett.} 118:144502

\bibitem[Michel et~al.(2018)Michel, Semin, Cazaubiel, Haudin, Humbert
  et~al.]{MichelPRF2018}
Michel G, Semin B, Cazaubiel A, Haudin F, Humbert T, et~al. 2018.
{Self-similar gravity wave spectra resulting from the modulation of bound
  waves}.
\textit{Phys. Rev. Fluids} 3:054801

\bibitem[Miles(1967)]{Miles1967}
Miles JW. 1967.
Surface-wave damping in closed basins.
\textit{Proc. R. Soc. A} 297:459--475

\bibitem[Miquel et~al.(2013)Miquel, Alexakis, Josserand \& Mordant]{Miquel2013}
Miquel B, Alexakis A, Josserand C, Mordant N. 2013.
Transition from wave turbulence to dynamical crumpling in vibrated elastic
  plates.
\textit{Phys. Rev. Lett.} 111:054302

\bibitem[Miquel et~al.(2014)Miquel, Alexakis \& Mordant]{Miquel2014}
Miquel B, Alexakis A, Mordant N. 2014.
Role of dissipation in flexural wave turbulence: from experimental spectrum to
  kolmogorov-zakharov spectrum.
\textit{Phys. Rev. E} 89:062925

\bibitem[Moisy et~al.(2009)Moisy, Rabaud \& Salsac]{Moisy2009}
Moisy F, Rabaud M, Salsac K. 2009.
{A Synthetic Schlieren method for the measurement of the topography of a liquid
  interface}.
\textit{Exp. Fluids} 46:1021--1036

\bibitem[Monsalve et~al.(2020)Monsalve, Brunet, Gallet \&
  Cortet]{MonsalvePRL2020}
Monsalve E, Brunet M, Gallet B, Cortet PP. 2020.
{Quantitative Experimental Observation of Weak Inertial-Wave Turbulence}.
\textit{Phys. Rev. Lett.} 125:254502

\bibitem[Nazarenko(2006)]{Nazarenko06}
Nazarenko S. 2006.
{Sandpile behaviour in discrete water-wave turbulence}.
\textit{J. Stat. Mech.} :L02002

\bibitem[Nazarenko(2011)]{NazarenkoBook2011}
Nazarenko S. 2011.
\em{Wave Turbulence}.
Springer-Verlag, Berlin

\bibitem[Nazarenko \& Lukaschuk(2016)]{NazarenkoRev2016}
Nazarenko S, Lukaschuk S. 2016.
{Wave Turbulence on Water Surface}.
\textit{Annu. Rev. of Cond. Mat. Phys.} 7:61

\bibitem[Nazarenko et~al.(2010)Nazarenko, Lukaschuk, McLelland \&
  Denissenko]{NazarenkoJFM2010}
Nazarenko S, Lukaschuk S, McLelland S, Denissenko P. 2010.
{Statistics of surface gravity wave turbulence in the space and time domains}.
\textit{J. Fluids Mech.} 642:395

\bibitem[Newell et~al.(2001)Newell, Nazarenko \& Biven]{Newell2001}
Newell AC, Nazarenko S, Biven L. 2001.
{Wave turbulence and intermittency}.
\textit{Physica D: Nonlinear Phenomena} 152-153:520--550

\bibitem[Newell \& Rumpf(2011)]{Newell2011}
Newell AC, Rumpf B. 2011.
{Wave Turbulence}.
\textit{Annu. Rev. Fluid Mech.} 43:59

\bibitem[Newell \& Zakharov(1992)]{NewellPRL92}
Newell AC, Zakharov VE. 1992.
{Rough sea foam}.
\textit{Phys. Rev. Lett.} 69:1149

\bibitem[Pan \& Yue(2014)]{PanPRL2014}
Pan Y, Yue DKP. 2014.
Direct numerical investigation of turbulence of capillary waves.
\textit{Phys. Rev. Lett.} 113:094501

\bibitem[Pan \& Yue(2015)]{PanJFM2015}
Pan Y, Yue DKP. 2015.
{Decaying capillary wave turbulence under broad-scale dissipation}.
\textit{J. Fluid. Mech.} 780:R1--1

\bibitem[Pan \& Yue(2017)]{PanYueJFM2017}
Pan Y, Yue DKP. 2017.
{Understanding discrete capillary-wave turbulence using a quasi-resonant
  kinetic equation}.
\textit{J. Fluid. Mech.} 816:R1--1

\bibitem[Peters(1985)]{Peters1985}
Peters F. 1985.
{Schlieren interferometry applied to a gravity wave in a density-stratified
  liquid}.
\textit{Exp Fluids} 3:261--269

\bibitem[Phillips(1958)]{PhillipsJFM1958}
Phillips OM. 1958.
{The equilibrium range in the spectrum of wind-generated waves}.
\textit{J. Fluid. Mech.} 4:426--433

\bibitem[Phillips(1960)]{PhillipsJFM1960}
Phillips OM. 1960.
{On the dynamics of unsteady gravity waves of finite amplitude. Part 1. The
  elementary interactions}.
\textit{J. Fluid. Mech.} 9:193--217

\bibitem[Prasad(2000)]{Prasad2000}
Prasad AK. 2000.
{Stereoscopic particle image velocimetry}.
\textit{Exp. Fluids} 29:103--116

\bibitem[Przadka et~al.(2012)Przadka, Cabane, Pagneux, Maurel \&
  Petitjeans]{Przadka2012}
Przadka A, Cabane B, Pagneux V, Maurel A, Petitjeans P. 2012.
{Fourier transform profilometry for water waves: How to achieve clean water
  attenuation with diffusive reflection at the water surface?}
\textit{Exp. Fluids} 52(2):519--527

\bibitem[Punzmann et~al.(2009)Punzmann, Shats \& Xia]{PunzmannPRL2009}
Punzmann H, Shats MG, Xia H. 2009.
{Phase Randomization of Three-Wave Interactions in Capillary Waves}.
\textit{Phys. Rev. Lett.} 103:064502

\bibitem[Pushkarev \& Zakharov(1996)]{Pushkarev1996}
Pushkarev AN, Zakharov VE. 1996.
{Turbulence of Capillary Waves}.
\textit{Phys. Rev. Lett.} 76:3320

\bibitem[Pushkarev \& Zakharov(2000)]{Pushkarev2000}
Pushkarev AN, Zakharov VE. 2000.
{Turbulence of capillary waves -- theory and numerical simulation}.
\textit{Physica D} 135:98

\bibitem[Redor et~al.(2019)Redor, Barth\'elemy, Michallet, Onorato \&
  Mordant]{RedorPRL2019}
Redor I, Barth\'elemy E, Michallet H, Onorato M, Mordant N. 2019.
Experimental evidence of a hydrodynamic soliton gas.
\textit{Phys. Rev. Lett.} 122:214502

\bibitem[Redor et~al.(2020)Redor, Barth{\'e}lemy, Mordant \&
  Michallet]{Redor2020}
Redor I, Barth{\'e}lemy E, Mordant N, Michallet H. 2020.
{Analysis of soliton gas with large-scale video-based wave measurements}.
\textit{Exp Fluids} 61:216

\bibitem[Ricard \& Falcon(2021{\natexlab{a}})]{Ricard2021}
Ricard G, Falcon E. 2021{\natexlab{a}}.
Experimental quasi-1{D} wave turbulence on the surface of a fluid.
Submitted to Europhys. Lett. (EPL)

\bibitem[Ricard \& Falcon(2021{\natexlab{b}})]{Ricard2021b}
Ricard G, Falcon E. 2021{\natexlab{b}}.
From dispersive to nondispersive collinear gravity-capillary wave turbulence.
In preparation

\bibitem[Savaro et~al.(2020)Savaro, Campagne, Linares, Augier, Sommeria
  et~al.]{SavaroPRF2020}
Savaro C, Campagne A, Linares MC, Augier P, Sommeria J, et~al. 2020.
{Generation of weakly nonlinear turbulence of internal gravity waves in the
  Coriolis facility}.
\textit{Phys. Rev. Fluids} 5:073801

\bibitem[Simmons(1969)]{Simmons1969}
Simmons WF. 1969.
{A variational method for week resonant wave interactions}.
\textit{Proc. Roy. Soc. A} 309:551--575

\bibitem[Snouck et~al.(2009)Snouck, Westra \& van~de Water]{Snouck2009}
Snouck D, Westra MT, van~de Water W. 2009.
{Turbulent parametric surface waves}.
\textit{Physics of Fluids} 21:025102

\bibitem[Suret et~al.(2020)Suret, Tikan, Bonnefoy, Copie, Ducrozet
  et~al.]{SuretPRL2020}
Suret P, Tikan A, Bonnefoy F, Copie F, Ducrozet G, et~al. 2020.
Nonlinear spectral synthesis of soliton gas in deep-water surface gravity
  waves.
\textit{Phys. Rev. Lett.} 125:264101

\bibitem[Takeda \& Mutoh(1983)]{TakedaAO1983}
Takeda M, Mutoh K. 1983.
{Fourier transform profilometry for the automatic measurement of 3-D object
  shapes}.
\textit{Appl. Opt.} 22:3977–3982

\bibitem[Tomita(1989)]{Tomita89}
Tomita H. 1989.
Theoretical and experimental investigations of interaction among deep-water
  gravity waves.
\textit{Rep. Ship Res. Inst.} 26:251

\bibitem[Turney et~al.(2009)Turney, Anderer \& Banerjee]{Turney2009}
Turney DE, Anderer A, Banerjee S. 2009.
A method for three-dimensional interfacial particle image velocimetry
  (3d-{IPIV}) of an air{\textendash}water interface.
\textit{Meas. Sci. Technol.} 20:045403

\bibitem[van Dorn(1966)]{VanDorn}
van Dorn WG. 1966.
{Boundary dissipation of oscillatory waves}.
\textit{J. Fluids Mech.} 24:769

\bibitem[Vedenov(1967)]{Vedenov1967}
Vedenov AA. 1967.
{Theory of a Weakly Turbulent Plasma}. In \textit{Reviews of Plasma Physics,
  Vol. 3}, ed. MA~Leontovich. Boston: Springer,  229--276

\bibitem[Whitham(1999)]{Whitham}
Whitham GB. 1999.
\em{Linear and Nonlinear Waves}.
Wiley-Interscience, Hoboken, NJ

\bibitem[Wright et~al.(1997)Wright, Budakian, Pine \& Putterman]{Wright1997}
Wright WB, Budakian R, Pine DJ, Putterman SJ. 1997.
Imaging of intermittency in ripple-wave turbulence.
\textit{Science} 278:1609

\bibitem[Wright et~al.(1996)Wright, Budakian \& Putterman]{Wright1996}
Wright WB, Budakian R, Putterman SJ. 1996.
Diffusing light photography of fully developed isotropic ripple turbulence.
\textit{Phys. Rev. Lett.} 76:4528

\bibitem[Xia et~al.(2010)Xia, Shats \& Punzmann]{XiaEPL2010}
Xia H, Shats M, Punzmann H. 2010.
{Modulation instability and capillary wave turbulence}.
\textit{Europhys. Lett.} 91:14002

\bibitem[Zakharov(1971)]{Zakharov1971}
Zakharov VE. 1971.
Kinetic equation for solitons.
\textit{JETP} 33:538--541

\bibitem[Zakharov(2010)]{Zakharov2010}
Zakharov VE. 2010.
{Energy balance in a wind-driven sea.}
\textit{Phys. Scr.} T142:014052

\bibitem[Zakharov et~al.(2019)Zakharov, Badulin, Geogjaev \&
  Pushkarev]{Zakharov2019}
Zakharov VE, Badulin SI, Geogjaev VV, Pushkarev AN. 2019.
Weak-turbulent theory of wind-driven sea.
\textit{Earth and Space Sci.} 6:540--556

\bibitem[Zakharov \& Filonenko(1967{\natexlab{a}})]{Zakharov1967grav}
Zakharov VE, Filonenko NN. 1967{\natexlab{a}}.
{Energy spectrum for stochastic oscillations of the surface of a liquid}.
\textit{Sov. Phys. Dokl.} 11:881

\bibitem[Zakharov \& Filonenko(1967{\natexlab{b}})]{Zakharov1967}
Zakharov VE, Filonenko NN. 1967{\natexlab{b}}.
{Weak turbulence of capillary waves}.
\textit{J. Appl. Mech. Tech. Phys} 8:37

\bibitem[Zakharov \& Filonenko(1968)]{Zakharov1968}
Zakharov VE, Filonenko NN. 1968.
{Stability of periodic waves of finite amplitude on a surface of a deep fluid}.
\textit{J. Appl. Mech. Tech. Phys} 2:190--198

\bibitem[Zakharov et~al.(2005)Zakharov, Korotkevich, Pushkarev \&
  Dyachenko]{Zakharov2005}
Zakharov VE, Korotkevich A, Pushkarev AN, Dyachenko AI. 2005.
{Mesoscopic wave turbulence}.
\textit{JETP Lett.} 82:487

\bibitem[Zakharov et~al.(1992)Zakharov, L'vov \& Falkovich]{ZakharovBook1992}
Zakharov VE, L'vov V, Falkovich G. 1992.
\em{Kolmogorov Spectra of Turbulence}.
Springer-Verlag, Berlin

\bibitem[Zakharov \& Zaslavskii(1982)]{Zakharov82GravInv}
Zakharov VE, Zaslavskii MM. 1982.
{The kinetic equation and Kolmogorov spectra in the weak turbulence theory of
  wind waves}.
\textit{Izv. Atm. Ocean Phys.} 18:747

\bibitem[Zavadsky et~al.(2017)Zavadsky, Benetazzo \& Shemer]{ZavadskyPOF2017}
Zavadsky A, Benetazzo A, Shemer L. 2017.
{On the two-dimensional structure of short gravity waves in a wind wave tank}.
\textit{Phys. Fluids} 29:016601

\end{thebibliography}
\bibliographystyle{ar-style1}

\end{document}